\documentclass[aps,prd,nofootinbib]{revtex4}
\usepackage{epsfig}
\usepackage{amssymb,amsmath}                                                    
\usepackage[colorlinks=true,linkcolor=blue,citecolor=blue, breaklinks=true]{hyperref}

\begin{document}
\title{Model Independent Foreground Power Spectrum Estimation \\
using WMAP $5$-year Data}
\author{Tuhin Ghosh$^1$, Rajib Saha$^{1,2,3,4}$, Pankaj Jain$^4$ and Tarun Souradeep$^1$}
\affiliation{$^1$IUCAA, Post Bag 4, Ganeshkhind, Pune-411007, India. \\$^2$Jet Propulsion Laboratory, M/S 169-327, 4800 Oak Grove Drive, Pasadena, CA 91109, USA.\\$^3$California Institute of Technology, Pasadena, CA 91125, USA. \\$^4$Department of Physics, Indian Institute of Technology, Kanpur, U.P, 208016, India. }

\email{tuhin@iucaa.ernet.in;rajib@caltech.edu;pkjain@iitk.ac.in;tarun@iucaa.ernet.in}

\begin{abstract}
\begin{center}
\textbf{Abstract}
\end{center}
In this paper, we propose \& implement on WMAP $5$-year data, a model independent approach of  foreground power spectrum estimation for multifrequency observations of CMB experiments. Recently a model independent approach of CMB  power spectrum estimation was proposed by Saha
et al. 2006. This methodology demonstrates that CMB power spectrum can be reliably estimated solely from WMAP data without assuming any template models for the foreground components. In the current paper, we extend this work to estimate the galactic foreground power spectrum using the WMAP $5$ year maps following a self contained analysis. We apply the model independent method in harmonic basis to estimate the foreground power spectrum and frequency
dependence of combined foregrounds. We also study the behaviour of synchrotron spectral index variation over different regions of the sky. We compare our results with those obtained from MEM foreground maps which are formed in pixel space. We find that relative to our model independent estimates MEM maps overestimates the foreground power close to galactic plane and underestimates it at high latitudes.
\end{abstract}

\maketitle
\section{Introduction}
The study of Galactic foreground emission in the range from few MHz to few hundreds of GHz is very important for CMB observations. Characterising the foreground emission unravels the galactic physics. Its understanding leads to more reliable removal of foreground contamination from the CMB temperature or polarization anisotropy maps. Diffuse galactic foregrounds consist of three main components : dust, free-free and synchrotron emission. Recent literature also report evidence of other possible contaminating components like spinning dust, hard synchrotron and galactic haze at WMAP frequencies \cite{fink1,fink2,Bennett:2003ba,Bennett}. It is
important to measure the galactic foregrounds components and their spectral behaviour unbiased by prior expectations. The physical mechanism and spectral behaviour of three main foreground components are summarised below:

\begin{itemize}

\item Synchrotron emission arises when an  electron moves at a relativistic velocity along a magnetic field line. In terms of antenna temperature the frequency dependence of synchrotron emission can be written as,
\begin{equation}
{T_{s}}\propto \nu^{\beta_s}\,,
\end{equation}
where the spectral index, $\beta_s$, varies across the sky. Based on the WMAP data, WMAP team estimated $\beta_s\simeq -3.5$ at high latitudes and $\beta_s\simeq -2.5$ close to star forming regions near the galactic plane.

\item Free-free emission arises when free electrons passing through the hot interstellar medium  are deflected and slowed down by ionized atoms mostly protons. The change of kinetic energy due to deflection is converted into free-free emission. The frequency dependence in
terms of antenna temperature can be written as,
\begin{equation}
{T_{f}}\propto \nu^{\beta_{f}}\,,
\end{equation}
where the index $\beta_f$ is flatter compared to $\beta_s$  and can be approximated as $\beta_f=-2.14$ in the WMAP frequency range.

\item Dust emission occurs when dust grains in the interstellar medium heated by the photon flux from the stars  seek thermal equilibrium by emitting in infra-red or far infra-red range of the spectrum. Dust emission can be well modeled by grey body spectrum and the frequency dependence is given by,
\begin{equation}
{T_d}\propto \frac{\nu^{{\beta_d}+1}}{exp(h\nu /KT_{dust})-1} \propto \nu^{\beta_{d}}\quad (h\nu<< KT_{dust})\,.
\end{equation}
\end{itemize}

Different techniques, such as, Maximum Entropy Method (MEM)\cite{Hinshaw:2006ia, Bennett, Gold:2008kp},  Correlated Component Analysis (CCA) \cite{bonaldi:07}, have been studied in the literature for modelling individual foreground components in pixel space. However, all these methods require foreground models in terms of templates, e.g., $408$MHz all sky synchrotron emission map (Haslam et al. 1982) \cite{haslam}, dust map at $94$GHz by Finkbeiner, Davis \& Schlegel (FDS99) \cite{fds} and Finkbeiner Halpha maps \cite{fink1}. While modeling foreground components these methods rely on extrapolation of the external templates to WMAP frequencies. There is always an uncertainty in extrapolating low frequency or high frequency template to WMAP frequencies \cite{lopez}. In this paper, we discuss about a model independent estimation of foreground behaviour without using any  information from external template. 

A model independent foreground removal method from the multifrequency CMB data was first proposed by Tegmark \& Efstathiou 1996, and was implemented on the WMAP $1$-year data by  Tegmark et al. 2003. This method was extended by \cite{saha1,saha2,tarun} to estimate CMB power spectrum by  cross-correlating cleaned  maps with independent detector noise.  We extend this method to get the spatial and spectral distributions of diffuse foreground power spectrum. 
The main advantage of this method is that it does not require any  assumptions about the foreground emission. It is based only on the fact that CMB anisotropy is independent of frequency in  thermodynamic temperature unit whereas foregrounds have a frequency dependence. In this paper, we  motivate ourselves  to estimate power spectrum of the composite emission due to all diffuse foreground  components in a self-contained manner using WMAP data only. However, we do not address the problem to present power spectrum due to individual foreground components. The method  does not require extrapolation of foreground templates from outside frequency or fitting any foreground template. Amongst other advantages are, it is computationally fast and unlike MEM or CCA our composite foreground maps are not limited by the lowest resolution frequency band. By extending the method we provide a map of the variation of the synchrotron spectral index over different regions of the sky.

The rest of the paper is organized as follows. Section II  briefly describes the model independent CMB power spectrum estimation methodology. Section III describes in detail the model independent foreground power spectrum estimation method. Section IV is dedicated to the estimation of synchrotron spectral index over different regions of the sky. Finally we conclude in Section V.

\section{Review of the CMB Power Spectrum Estimation}
The basic assumption of model independent CMB power spectrum estimation is that CMB contributes equally in all frequency channels (in terms of thermodynamic temperature). The temperature anisotropy at a frequency channel $i$ can be written as,
\begin{equation}
\label{eq:a}
 \Delta T^i(\hat{n})= \int \lbrace\Delta T^{C}(\hat{n})+ \Delta T^{F}_i(\hat{n})\rbrace B_i(\hat{n}.\hat{n}')\,d\hat{n}' + \Delta T^{N}_i(\hat{n}),
\end{equation}
where $\Delta T^{C}(\hat{n})$ and $\Delta T^{F}_i(\hat{n})$ are respectively the CMB and foreground components  and $\Delta T^{N}_i(\hat{n})$ is the noise component. Here $B_i(\hat{n}.\hat{n}'$) is the circularly symmetric beam denoting the smoothing of the map due to the finite resolution of the detector. The spherical harmonic transform of the map at the frequency channel $i$ can be written as,
\begin{equation*}
 a_{lm}^i =  B^i_l a_{lm}^C + B^i_l a_{lm}^F(i) + a_{lm}^N(i),
\end{equation*}
where the index $i$ runs from $1$ to $5$ for five frequency bands of WMAP. In our  analysis, we perform foreground removal on harmonic space and separately over different regions of the sky. We first partition the entire sky in nine different regions according to their level of foreground contamination as described in \cite{Tegmark:2003ve,saha1} and then perform foreground removal for each region iteratively.  Below we briefly describe the basic algorithm of the procedure. 

We define a cleaned map as a sum of linearly weighted  $5$  WMAP maps as follows,
\begin{equation}
 a_{lm}^{clean}=\sum_{i=1}^{5} W_l^{i}\frac{a_{lm}^{i}}{B_l^{i}},
\label{eq:b}
\end{equation}
where the weight factor depends on the frequency $i$ as well as on the multipole $l$. Since the frequency channels of WMAP are of different resolution, we deconvolve each map by the corresponding beam, $B_l^{i}$, prior to linear combination. The power spectrum of the cleaned map is given by, 
\begin{align*}
 C_l^{clean} &= \frac{1}{2l+1}\sum_{m=-l}^l a_{lm}^{clean} a_{lm}^{clean*}. \\
 &=  \frac{1}{2l+1}\sum_{m=-l}^l\hspace{0.2cm}\sum_{i=1}^{5} W_l^{i}\frac{a_{lm}^{i}}{B_l^{i}}\sum_{j=1}^{5} W_l^{j}\frac{a_{lm}^{j*}}{B_l^{j}}\hspace{0.2cm} .
\end{align*}
The above equation can be simplified and written as a matrix equation as,
\begin{align*}
C_l^{clean} &= \mathbf{W}_l \mathbf{C}_l \mathbf{W}_l^T\\
&= \mathbf{W}_l (\mathbf{C}_l^S+\mathbf{C}_l^R) \mathbf{W}_l^T\\
&= C_l^S \mathbf{W}_l\mathbf{e} \mathbf{e}^T \mathbf{W}_l^T + \mathbf{W}_l\mathbf{C}_l^R\mathbf{W}_l^T,
\end{align*}
where $\mathbf{e}$ = (1, 1, ..,1) is a column vector of five ones, `$\mathbf {C}^S_l$' and `$\mathbf {C}^R_l$' represent respectively the CMB and non-CMB ( i.e., foreground plus detector noise) covariance matrices. `${C}^S_l$' is the CMB power spectrum. To preserve the CMB power spectrum in the cleaned map we impose the condition, $\mathbf{W}_l\mathbf{e}$ = $\mathbf{e}^T\mathbf{W}_l^T = 1$. 
Hence we can re-express above equation as follows,
\begin{equation*}
 C_l^{clean} = C_l^S  + \mathbf{W}_l\mathbf{C}_l^R\mathbf{W}_l^T.
\end{equation*}
Since in the above equation, CMB power spectrum remains independent on weights, minimizing $C_l^{clean}$  minimizes the combined contamination coming from foregrounds and the detector noise without altering  the CMB power. Minimization of $C_l^{clean}=\mathbf{W}_l \mathbf{C}_l \mathbf{W}_l^T$ with the condition that $\mathbf{W}_l\mathbf{e}$ = $\mathbf{e}^T\mathbf{W}_l^T = 1$ is a standard Lagrangian multiplier problem and  has a well known solution,
\begin{equation}
 \mathbf{W}_l^T=\frac{\mathbf{C}_l^{-1}\mathbf{e}}{\mathbf{e}^T\mathbf{C}_l^{-1}\mathbf{e}}.
\end{equation}
Putting this $\mathbf{W}_l^T$ back in equation~\eqref{eq:b}, we get the cleaned CMB map. The cleaned map in fourier space can be written as,
\begin{equation}
\label{eq:c}
 a_{lm}^{clean}(i) =  a_{lm}^C(i) + a_{lm}^{RN}(i),
\end{equation}
where `RN' denotes the residual noise in the cleaned map.

\section{Estimation of Foreground Power Spectrum}
 The WMAP satellite has 10 differential assemblies distributed over  five different frequencies ranging from 23GHz to 94GHz. K and Ka band  have one differential assembly (DA) each. Q and V band have two DA's namely Q1, Q2, V1 and V2 respectively. W band has four DA's namely W1, W2, W3 and W4. For W band, we simply average the pairs of DA's in  the W band to form 6 DA maps - W12, W13, W14, W23, W24, W34 out of 4 DA's of W band. In our analysis, we smooth all the 5Yr WMAP maps to a common beam resolution of $1^0$. To estimate the cleaned CMB maps, we linearly combine the 4 DA's of different frequencies at time as described in Sec II and in details in \cite{saha1,saha2}. The various different 4 channel combination and 3 channel combination cleaned maps that can be obtained are listed in table~\ref{tab:1}.

\begin{table}[h!]
\begin{minipage}{2in}
\scriptsize
\begin{tabular}{|c|c|}
\hline
\multicolumn{2}{|c|}{\bf 4-channel combinations  ($n_c=4$)} \\
\hline
 & \\
(K,KA)+Q1+V1+W12=(C1,CA1)&(K,KA)+Q1+V2+W12=(C13,CA13)\\
(K,KA)+Q1+V1+W13=(C2,CA2)&(K,KA)+Q1+V2+W13=(C14,CA14)\\
(K,KA)+Q1+V1+W14=(C3,CA3)&(K,KA)+Q1+V2+W14=(C15,CA15)\\
(K,KA)+Q1+V1+W23=(C4,CA4)&(K,KA)+Q1+V2+W23=(C16,CA16)\\
(K,KA)+Q1+V1+W24=(C5,CA5)&(K,KA)+Q1+V2+W24=(C17,CA17)\\
(K,KA)+Q1+V1+W34=(C6,CA6)&(K,KA)+Q1+V2+W34=(C18,CA18)\\
(K,KA)+Q2+V2+W12=(C7,CA7)&(K,KA)+Q2+V1+W12=(C19,CA19)\\
(K,KA)+Q2+V2+W13=(C8,CA8)&(K,KA)+Q2+V1+W13=(C20,CA20)\\
(K,KA)+Q2+V2+W14=(C9,CA9)&(K,KA)+Q2+V1+W14=(C21,CA21)\\
(K,KA)+Q2+V2+W23=(C10,CA10)&(K,KA)+Q2+V1+W23=(C22,CA22)\\
(K,KA)+Q2+V2+W24=(C11,CA11)&(K,KA)+Q2+V1+W24=(C23,CA23)\\
(K,KA)+Q2+V2+W34=(C12,CA12)&(K,KA)+Q2+V1+W34=(C24,CA24)\\
                           &                           \\
\hline
\end{tabular}
\end{minipage}
\hspace{5cm}
\begin{minipage}{2in}
\scriptsize
\begin{tabular}{|c|c|}
\hline
\multicolumn{2}{|c|}{\bf 3-channel combinations  ($n_c=3$)} \\
\hline
                   &                            \\
Q1+V1+W12=C1 & Q1+V2+W12=C13\\
Q1+V1+W13=C2 & Q1+V2+W13=C14\\
Q1+V1+W14=C3 & Q1+V2+W14=C15\\
Q1+V1+W23=C4 & Q1+V2+W23=C16\\
Q1+V1+W24=C5 & Q1+V2+W24=C17\\
Q1+V1+W34=C6 & Q1+V2+W34=C18\\
Q2+V2+W12=C7 & Q2+V1+W12=C19\\
Q2+V2+W13=C8 & Q2+V1+W13=C20\\
Q2+V2+W14=C9 & Q2+V1+W14=C21\\
Q2+V2+W23=C10 & Q2+V1+W23=C22\\
Q2+V2+W24=C11 & Q2+V1+W24=C23\\
Q2+V2+W34=C12 & Q2+V1+W34=C24\\
                            &                      \\
\hline
\end{tabular}
\end{minipage}
 \caption{The table shows 48 different combinations of the DA maps used in our 4 channel cleaning method and list of the 24 possible combinations in the 3 channel cleaning method~\cite{saha1,saha2}.}
\label{tab:1}
\end{table}

The spherical harmonic transform of the WMAP maps at each frequency channel referred by index \textit{i}, smoothed at $1^0$ beam resolution can be written as,
\begin{equation*}
 a_{lm}^{i} =  B_l a_{lm}^C + B_l a_{lm}^F(i) + a_{lm}^N(i).
\end{equation*}
The one degree smooth cleaned map can be written as,
\begin{equation}
 a_{lm}^{clean} =  B_l a_{lm}^C + B_l a_{lm}^{RN},
\end{equation}
where `$RN$' denotes the residual noise in the single cleaned map obtained by our model independent analysis and $B_l$ is the fourier transform of the beam at one degree resolution. To obtain the foreground power spectrum, we subtract the cleaned maps from the DA maps
obtained as described earlier. The spherical harmonic transform of Di's (see table~\ref{tab:2}) which is obtained after subtracting the cleaned cmb maps at each frequency channel can now be written as,
\begin{align}
 a_{lm}^{i} -  a_{lm}^{clean} &=  B_l a_{lm}^F(i) + ( a_{lm}^N(i) - B_l a_{lm}^{RN}).\notag\\
&=  B_l a_{lm}^F(i) + a_{lm}^{N'}(i).\notag
\end{align}
To estimate the foreground power spectrum at $1^0$ beam resolution, we remove the noise bias by cross correlating pairs of CMB free maps which have no common DA/detector in the cleaned maps involved. Assuming there is no cross correlation between the foreground and noise and
independent noises for two different detector, we can write the $1^0$ beam smoothed foreground power spectrum $C_l^F$ as,
\begin{align}
\langle(a_{lm}^{Total}(i) -  a_{lm}^{clean}&(i))(a_{lm}^{Total*}(j) -  a_{lm}^{clean*}(j))\rangle\notag\\
& =\langle B_l^2 a_{lm}^F(i)a_{lm}^{F*}(j)\rangle + \langle a_{lm}^{N'}(i) a_{lm}^{N'*}(j)\rangle.\notag\\
&=C_l^F.\label{eq:d}
\end{align}

\begin{table}[h!]
\scriptsize
\begin{tabular}{|c|c|cc|}
\hline
\multicolumn{2}{|c|}{\bf CMB subtracted WMAP Q band} & \multicolumn{2}{|c|}{\bf Cross combinations}\\
\hline
 &  & & \\
Q1-C[(K,KA)+Q1+V1+W12]=(D01,DA01)& Q2-C[(K,KA)+Q2+V2+W34]=(D13,DA13) & D01 $\otimes$ DA13 & \quad DA01 $\otimes$ D13\\
Q1-C[(K,KA)+Q1+V1+W13]=(D02,DA02)& Q2-C[(K,KA)+Q2+V2+W24]=(D14,DA14) & D02 $\otimes$ DA14 & \quad DA02 $\otimes$ D14\\
Q1-C[(K,KA)+Q1+V1+W14]=(D03,DA03)& Q2-C[(K,KA)+Q2+V2+W23]=(D15,DA15) & D03 $\otimes$ DA15 & \quad DA03 $\otimes$ D15\\
Q1-C[(K,KA)+Q1+V1+W23]=(D04,DA04)& Q2-C[(K,KA)+Q2+V2+W14]=(D16,DA16) & D04 $\otimes$ DA16 & \quad DA04 $\otimes$ D16\\
Q1-C[(K,KA)+Q1+V1+W24]=(D05,DA05)& Q2-C[(K,KA)+Q2+V2+W13]=(D17,DA17) & D05 $\otimes$ DA17 & \quad DA05 $\otimes$ D17\\
Q1-C[(K,KA)+Q1+V1+W34]=(D06,DA06)& Q2-C[(K,KA)+Q2+V2+W12]=(D18,DA18) & D06 $\otimes$ DA18 & \quad DA06 $\otimes$ D18\\
Q1-C[(K,KA)+Q1+V2+W12]=(D07,DA07)& Q2-C[(K,KA)+Q2+V1+W34]=(D19,DA19) & D07 $\otimes$ DA19 & \quad DA07 $\otimes$ D19\\
Q1-C[(K,KA)+Q1+V2+W13]=(D08,DA08)& Q2-C[(K,KA)+Q2+V1+W24]=(D20,DA20) & D08 $\otimes$ DA20 & \quad DA08 $\otimes$ D20\\
Q1-C[(K,KA)+Q1+V2+W14]=(D09,DA09)& Q2-C[(K,KA)+Q2+V1+W23]=(D21,DA21) & D09 $\otimes$ DA21 & \quad DA09 $\otimes$ D21\\
Q1-C[(K,KA)+Q1+V2+W23]=(D10,DA10)& Q2-C[(K,KA)+Q2+V1+W14]=(D22,DA22) & D10 $\otimes$ DA22 & \quad DA10 $\otimes$ D22\\
Q1-C[(K,KA)+Q1+V2+W24]=(D11,DA11)& Q2-C[(K,KA)+Q2+V1+W13]=(D23,DA23) & D11 $\otimes$ DA23 & \quad DA11 $\otimes$ D23\\
Q1-C[(K,KA)+Q1+V2+W34]=(D12,DA12)& Q2-C[(K,KA)+Q2+V1+W12]=(D24,DA24) & D12 $\otimes$ DA24 & \quad DA12 $\otimes$ D24\\
                           &              & &             \\
\hline
\end{tabular}
 \caption{The table shows 48 possible combinations of the CMB subtracted WMAP Q band maps and 24 cross combinations to get rid of noise.}
\label{tab:2}
\end{table}

Here we explain the steps followed to obtain the foreground power spectrum for Q band where the number of DA's are more than one. Similar steps are also followed for V and W band.
\begin{itemize}
\item We smooth all the WMAP DA maps and the cleaned maps to one degree beam resolution.

\item Take the cleaned map C1 from 4-channel combinations as given in table~\ref{tab:1} and subtract it from Q1 map. And similarly take the cleaned map C12 from 4-channel combinations and subtract it from Q2 map. We label the CMB subtracted maps as Di as given in table~\ref{tab:2}.

\item Cross correlate the D01(Q1-C1) with the D13(Q2-C12) as discuss in table~\ref{tab:2} in  equation~\eqref{eq:d} to get rid of noise term and obtain the foreground power spectrum.

\item Repeat the above steps for 24 possible combinations obtained by 4 channel cleaned maps given in table~\ref{tab:2} and while cross correlate we choose pairs of Di where detectors are not common. Obtain the mean foreground power spectrum.
 
\end{itemize}

Since, K and Ka band have only one DA, cross correlation isn't feasible. In these cases, estimate of contribution of noise bias is explicitly calculated from the number of observation and sensitivity of the detector and subtracted out. To avoid any kind of mixing, we subtract the 3 channel cleaned maps C1 from let say K band and cross correlated with another 3 channel cleaned map C12 (where the detector used are not common) subtracted K map. So, the equation~\eqref{eq:d} in case of one DA will become,
\begin{align}
\langle(a_{lm}^{Total}(i) -  a_{lm}^{clean}(i))&(a_{lm}^{Total*}(j) -  a_{lm}^{clean*}(j))\rangle \notag\\
&=\langle(B_l^2 a_{lm}^F(i)a_{lm}^{F*}(j))\rangle + \langle(a_{lm}^N(i)a_{lm}^{N*}(j))\rangle. \notag\\
\label{eq:e}
&=C_l^F + C_l^N.
\end{align}
\begin{table}[h!]
\scriptsize
\begin{tabular}{|c|c|c|}
\hline
\multicolumn{2}{|c|}{\bf CMB subtracted WMAP K maps} & Cross combinations\\
\hline
 &  & \\
K-C(Q1V1W12)=D01 & K-C(Q2V2W34)=D13 & D01 $\otimes$ D13 \\
K-C(Q1V1W13)=D02 & K-C(Q2V2W24)=D14 & D02 $\otimes$ D14 \\
K-C(Q1V1W14)=D03 & K-C(Q2V2W23)=D15 & D03 $\otimes$ D15 \\
K-C(Q1V1W23)=D04 & K-C(Q2V2W14)=D16 & D04 $\otimes$ D16 \\
K-C(Q1V1W24)=D05 & K-C(Q2V2W13)=D17 & D05 $\otimes$ D17 \\
K-C(Q1V1W34)=D06 & K-C(Q2V2W12)=D18 & D06 $\otimes$ D18 \\
K-C(Q1V2W12)=D07 & K-C(Q2V1W34)=D19 & D07 $\otimes$ D19 \\
K-C(Q1V2W13)=D08 & K-C(Q2V1W24)=D20 & D08 $\otimes$ D20 \\
K-C(Q1V2W14)=D09 & K-C(Q2V1W23)=D21 & D09 $\otimes$ D21 \\
K-C(Q1V2W23)=D10 & K-C(Q2V1W14)=D22 & D10 $\otimes$ D22 \\
K-C(Q1V2W24)=D11 & K-C(Q2V1W13)=D23 & D11 $\otimes$ D23 \\
K-C(Q1V2W34)=D12 & K-C(Q2V1W12)=D24 & D12 $\otimes$ D24 \\
\hline
\end{tabular}
\label{tab:3}
 \caption{The table shows 24 different combinations of the DA maps for K band and 12 cross combinations to get rid of noise. There is a corresponding set for the Ka band.}
\end{table}

The second term in right hand side $C_l^N$ of equation~\eqref{eq:e} can easily be estimated using the relation,
\begin{equation*}
C_l^N= \frac {B_l^2(1^0)}{B_l^2(K)}\frac{4\pi}{N_{pix}^2}\sum_{i=1}^{N_{pix}}\frac{\sigma 
_i^2}{N_{obs}}.
\end{equation*}
By substituting the value of $C_l^N$ back in equation~\eqref{eq:e} and subtracting it from the left hand side gives the $1^0$ beam smooth foreground power spectrum ($C_l^F$). In general, foreground power spectra are expressed in terms of antenna temperature which can easily be converted from thermodynamic temperature using the table~\ref{table1}. Henceforth, all the results and plots of $C_l^F$ is expressed in terms of antenna temperature.\\
To get the foreground power spectrum outside the KQ85 mask, we mask the difference map Di's maps with combine KQ85 and WMAP5 point source mask and then smooth to 1 degree before cross-correlating the Di's where no common DA/detector are present to get rid of noise. Additional smoothing of $1^0$ is applied outside the KQ85 mask to compare the foreground power spectrum with MEM maps. 

\begin{table}[h!]
\begin{tabular}{|c|c|c|c|}
\hline
Frequency& Conversion Factor & Mean $\sigma_0$ & Mean FWHM\\
(in GHz) & g($\nu$) & (in mK) & (in degrees)\\
\hline
23  & 0.9867 & 1.436 & 0.807\\
33  & 0.9723 & 1.470 & 0.624\\
41  & 0.9581 & 2.197 & 0.4775\\
61  & 0.9095 & 3.133 & 0.326\\
94  & 0.8012 & 6.538 & 0.2038\\
\hline
\end{tabular}
\caption{Conversion factor from thermodynamic to antenna temperature where $\Delta T_A= g(\nu)\Delta T$;
 $g(\nu)=[{x^2e^x}/{(e^x-1)^2}]$; $x$=${h\nu}/{k_BT_0}$; $T_0$=2.725K.}
\label{table1}
\end{table}

\begin{figure}[h!]
\begin{tabular}{cc}
\epsfig{file=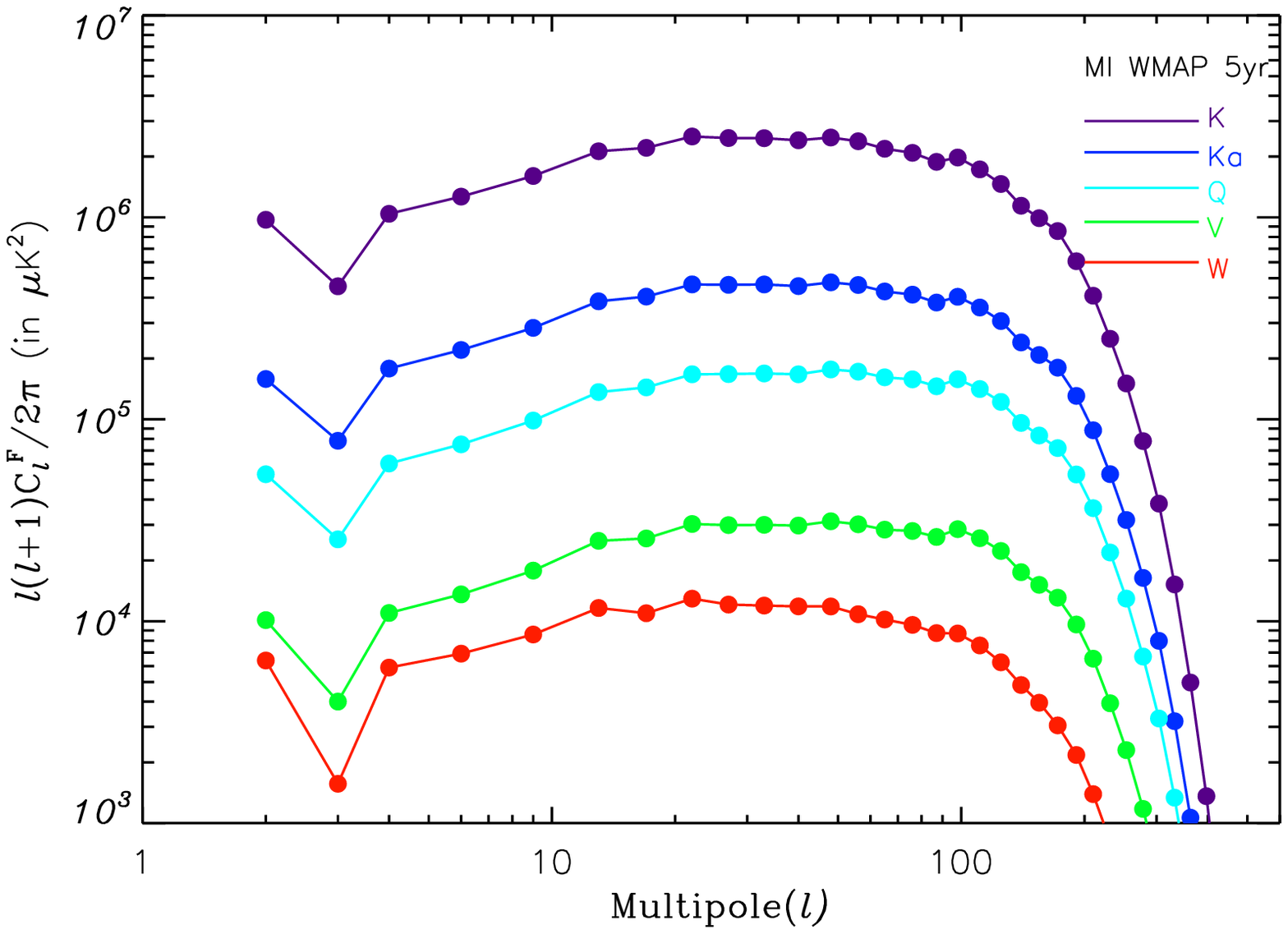,width=0.5\linewidth,angle=0,clip=}&
\epsfig{file=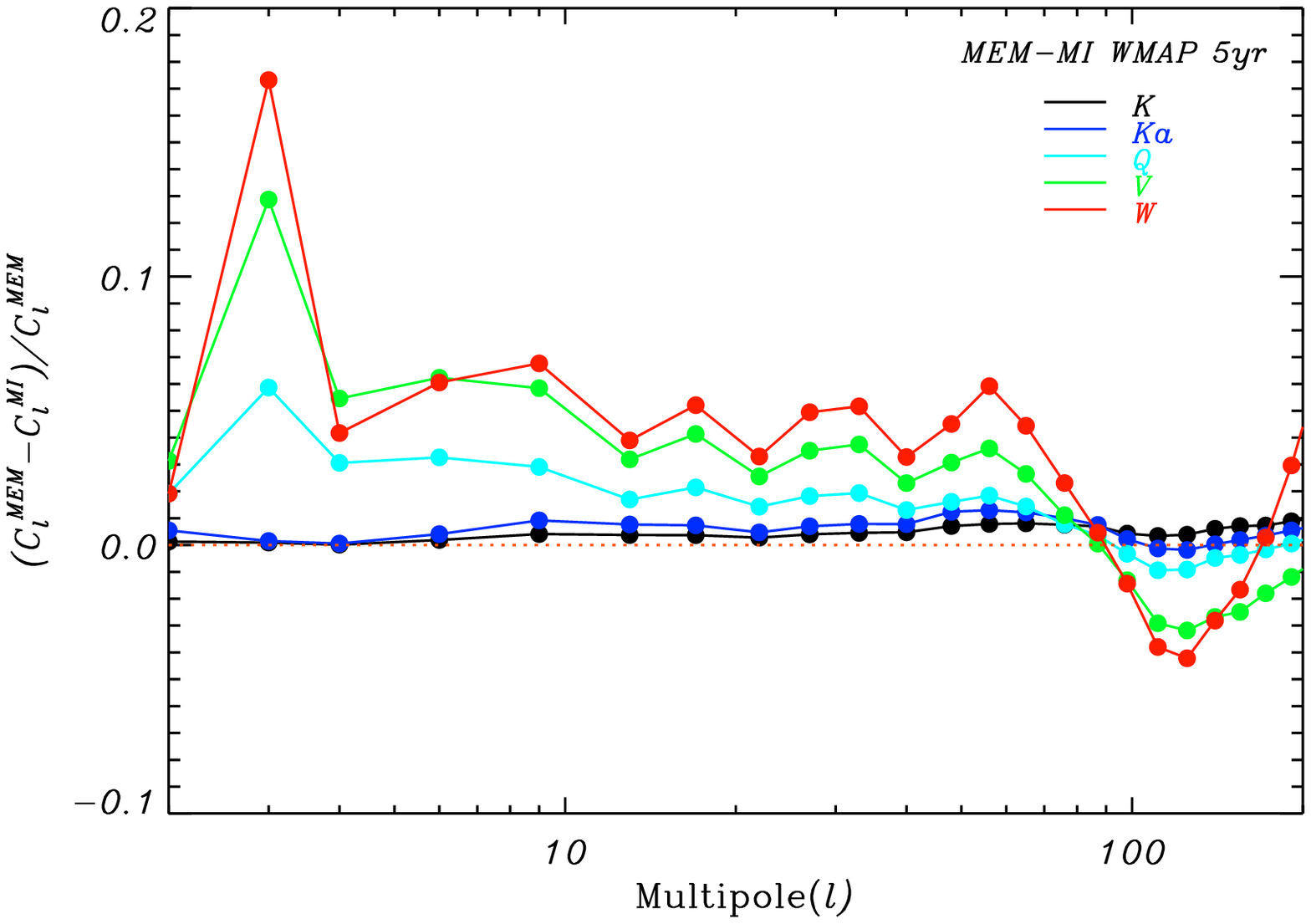,width=0.5\linewidth,angle=0,clip=}\\
\epsfig{file=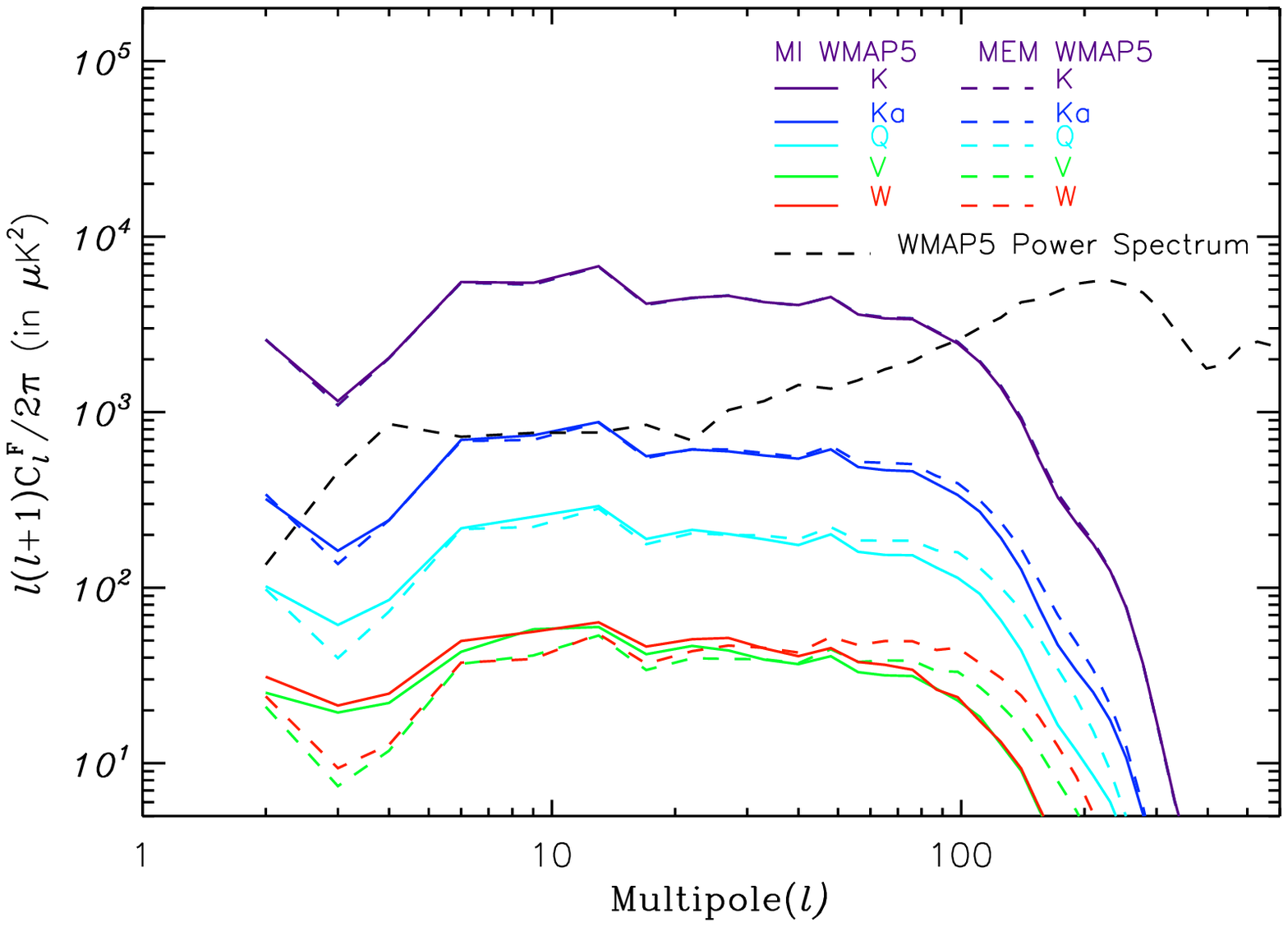,width=0.5\linewidth,angle=0,clip=}&
\epsfig{file=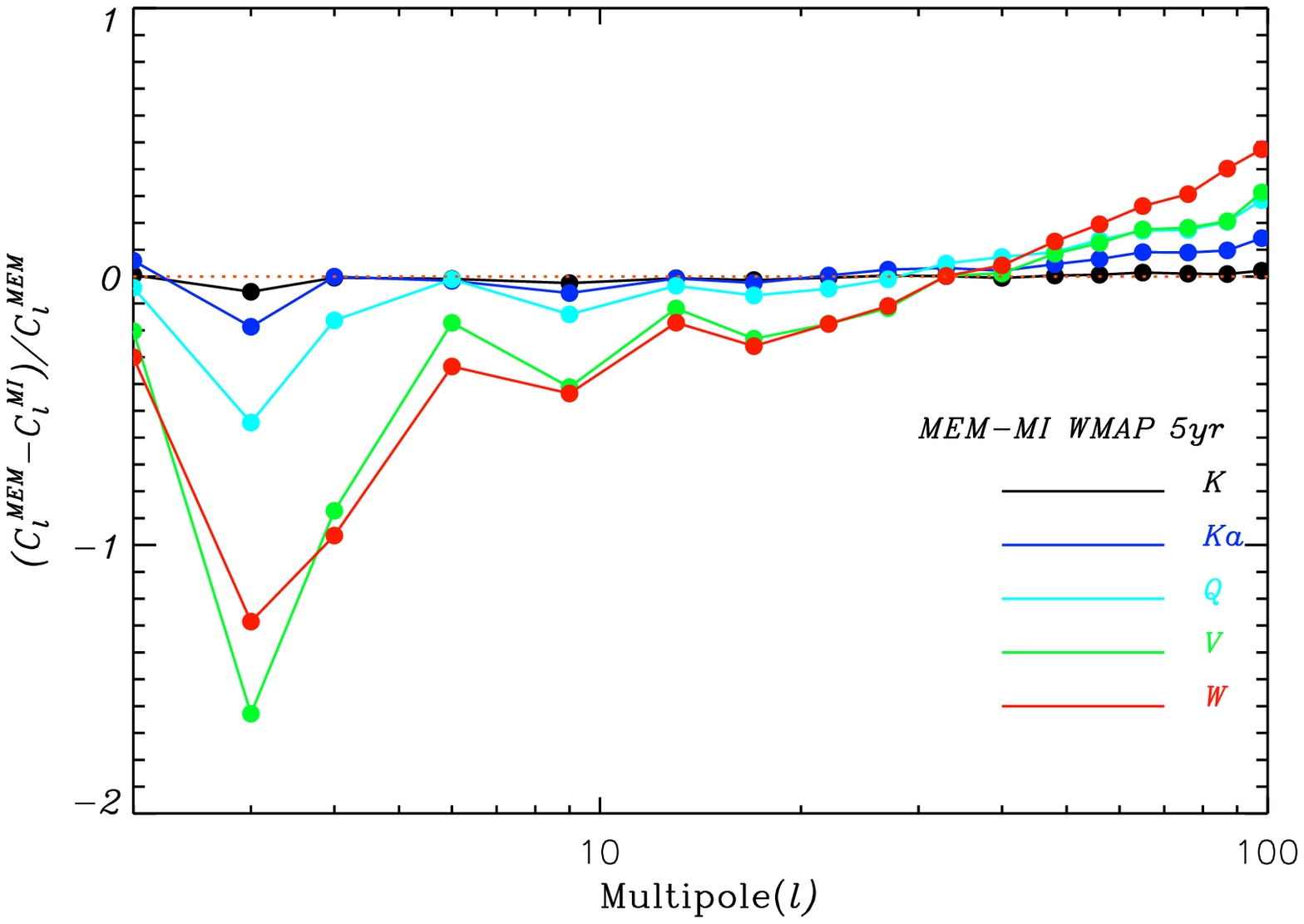,width=0.5\linewidth,angle=0,clip=}\\
\end{tabular}
\caption{\textit{Top Left Panel}: The model independent estimate of angular power spectrum for
combined foregrounds at WMAP frequency starting from K band to W band. In this case $C_l^F$ is smoothed by 1 degree beam and expressed in terms of antenna temperature. \textit{Top Right Panel}: Plot of relative difference in estimated power $({C_l^{MEM}-C_l^{MI}})/C_l^{MEM}$ with the multipole $l$, where `$MI$' stands for the foreground power spectrum obtained from model
independent analysis. Foreground power spectrum at K and Ka band is consistent with the WMAP Team. But for Q, V and W band we obtain slightly less power compared to MEM method. \textit{Bottom Left Panel}: The angular power spectrum of combined foregrounds outside KQ85 mask at WMAP frequencies starting from K band to W band expressed in antenna temperature. \textit{Bottom Right Panel}: Plot of relative power difference $({C_l^{MEM}-C_l^{MI}})/C_l^{MEM}$ with the multipole $l$ clearly shows that at low multipole MEM underestimates the foreground power outside KQ85 mask.}
\label{fig:1}
\end{figure}

We can calculate the r.m.s temperature, $\Delta T_{rms}$, of the foregrounds using the relation, 
\begin{equation*}
(\Delta T_{rms})^2 = \sum_{l=2}^{l_{max}} \frac{2l+1}{4\pi} C_l^F,
\end{equation*}
where $\Delta T_{rms}$ is expressed in terms of antenna temperature. We obtain the frequency dependent $\Delta T_{rms}$ for the five WMAP frequencies. We can model the rms foreground power in terms of three major galactic foregrounds -- synchrotron, free-free and dust emission. We fit the $\Delta T_{rms}$ at the five frequencies to obtain the full sky synchrotron spectral index,
\begin{align}
 \Delta T_{rms} &= A_s\nu^{\beta_s}+A_{f}\nu^{\beta_{f}}+ A_d\nu^{\beta_d}.
\end{align}
where $\beta_d=1.8$, $\beta_{f}=-2.14$ are taken as a constant parameters.

\begin{table}[h!]
\begin{tabular}{|c|c|c|}
\hline
Frequency & RMS Temperature & RMS Temperature\\
\textit{(in GHz)}  & using MI Analysis & using MEM maps.\\
  & \textit{(in mK)} & \textit{(in mK)}\\
\hline
23  & 2.795 & 2.801\\
33  & 1.210 & 1.213\\
41  & 0.731 & 0.736\\
61  & 0.310 & 0.314\\
94  & 0.198 & 0.201\\
\hline
\end{tabular}
\caption{Comparison of rms foreground power obtained from Model Independent analysis and MEM method. The rms foreground power matches closely with MEM method with a minor deficient seen in the Model Independent case.}
\end{table}

\begin{figure}[h!]
\mbox{\epsfig{file=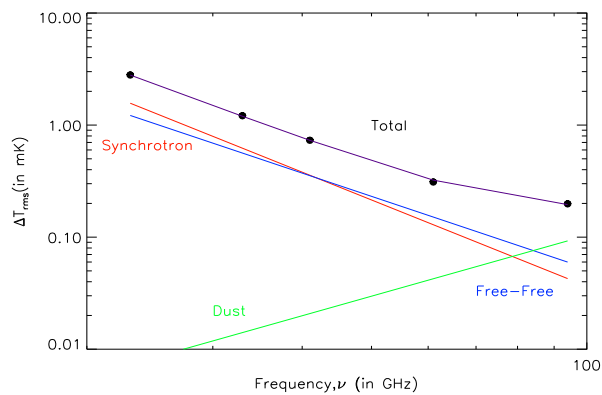,width=3.25in,height=2.2in,angle=0}}
\label{fig:figure2}
\caption{The total foreground emission (black dots) spectra obtained using model independent analysis compared to sum of foreground components (deep blue line connecting black dots) over the full sky. The average synchrotron spectrum from K to W band is -2.6 assuming the free-free spectral index $\beta_{f}=-2.14$ and dust spectral index  $\beta_d=1.8$ are constant parameters. The average synchrotron spectral index from K-Ka band and Ka-Q band are -3.0 and -2.91 respectively.} 
\end{figure}

As a consistency check, we compare our results with that obtained by the WMAP team using Maximum Entropy method (MEM) analysis. We find that our methods detects marginally lesser foreground power spectrum for Q to W band as compared to WMAP MEM method over the full sky. K and Ka band is quite consistent as shown in figure(\ref{fig:1}). The excess power in WMAP MEM
method comes from the low multipoles. To estimate the foreground power outside KQ85 region, MEM maps are first mask with KQ85 mask and then smoothed to $1^0$ beam resolution to get  foreground power spectrum of effective smoothing of $1.414^0$. We found that MEM method underestimates the foreground power at low multipoles outside KQ85 mask. The advantage of model independent analysis is that foreground power spectrum is not resolution limited. For each band, we can estimate the foreground power spectrum up to multipoles limited by the beam of that band.

\begin{figure}[h!]
\begin{tabular}{ccc}
\epsfig{file=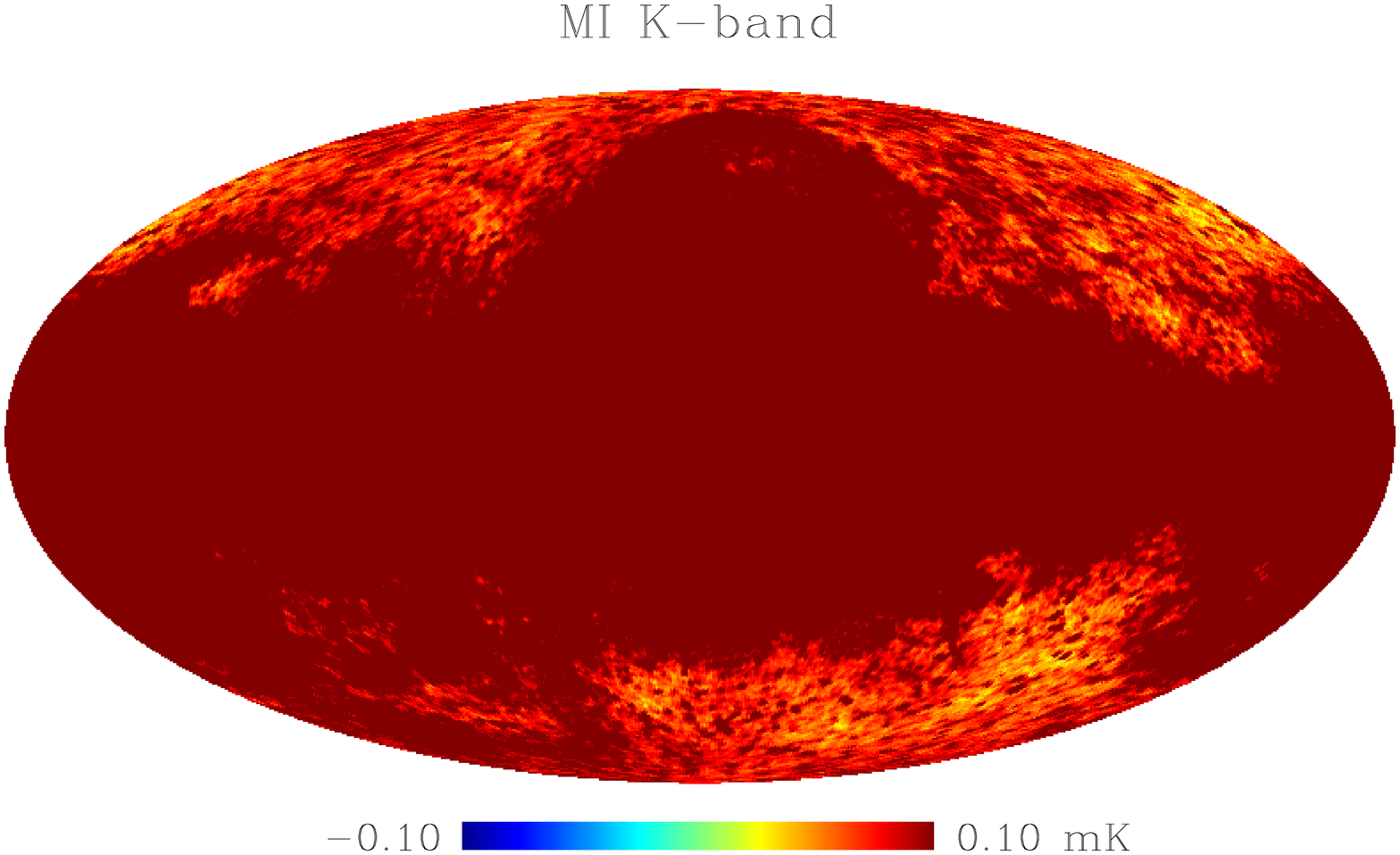,width=0.2\linewidth,angle=90,clip=} &
\epsfig{file=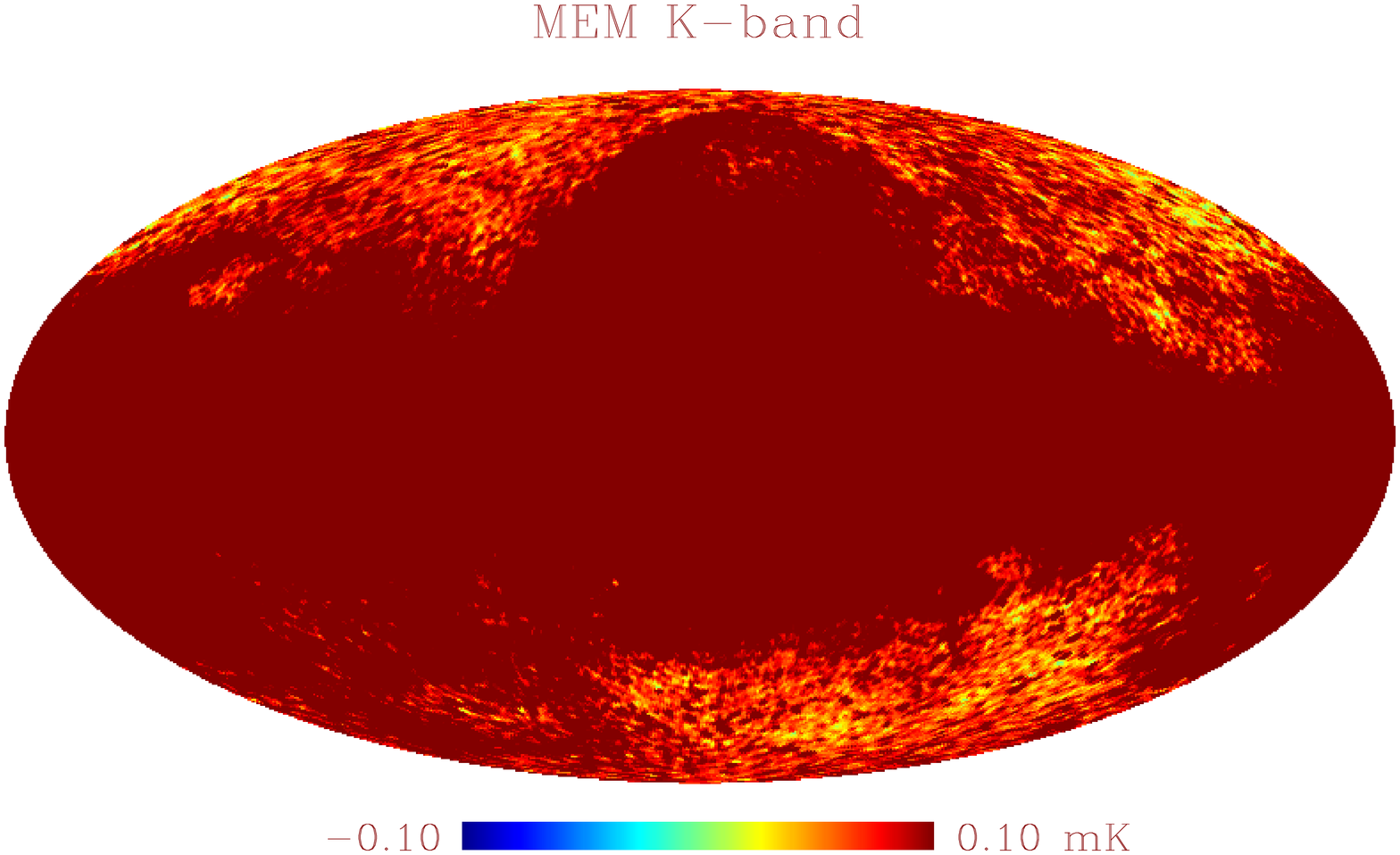,width=0.2\linewidth,angle=90,clip=} &
\epsfig{file=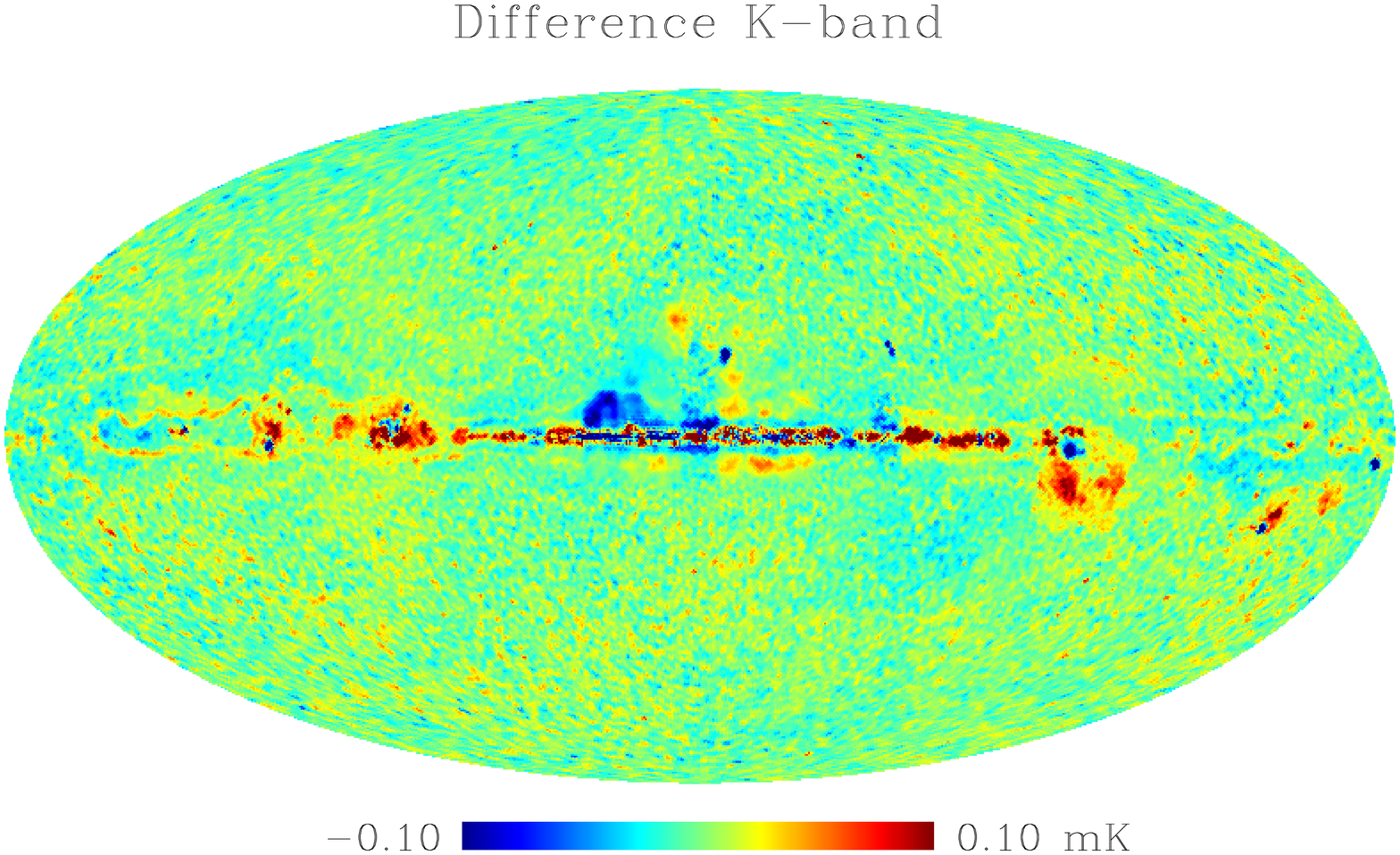,width=0.2\linewidth,angle=90,clip=} \\
\epsfig{file=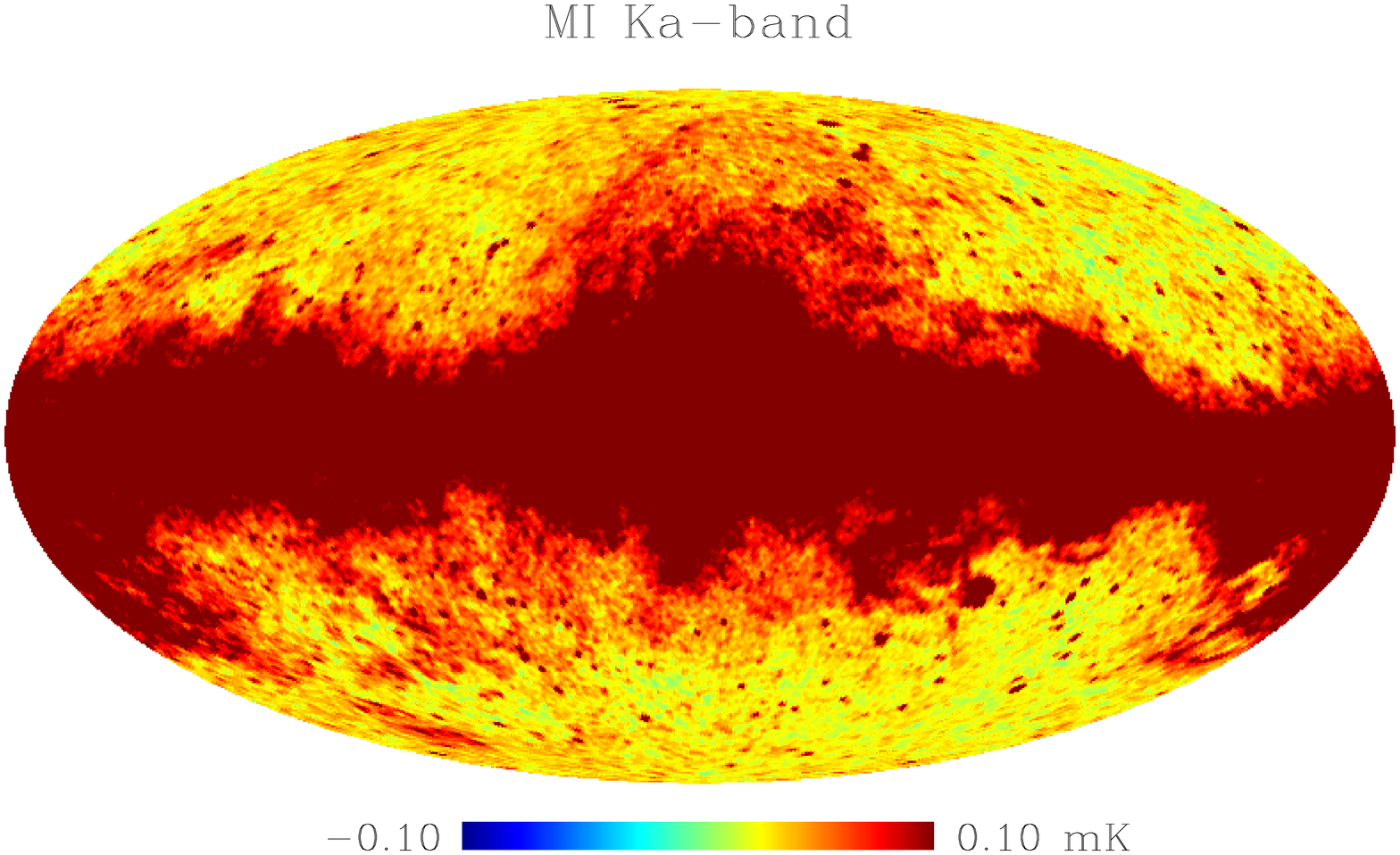,width=0.2\linewidth,angle=90,clip=} &
\epsfig{file=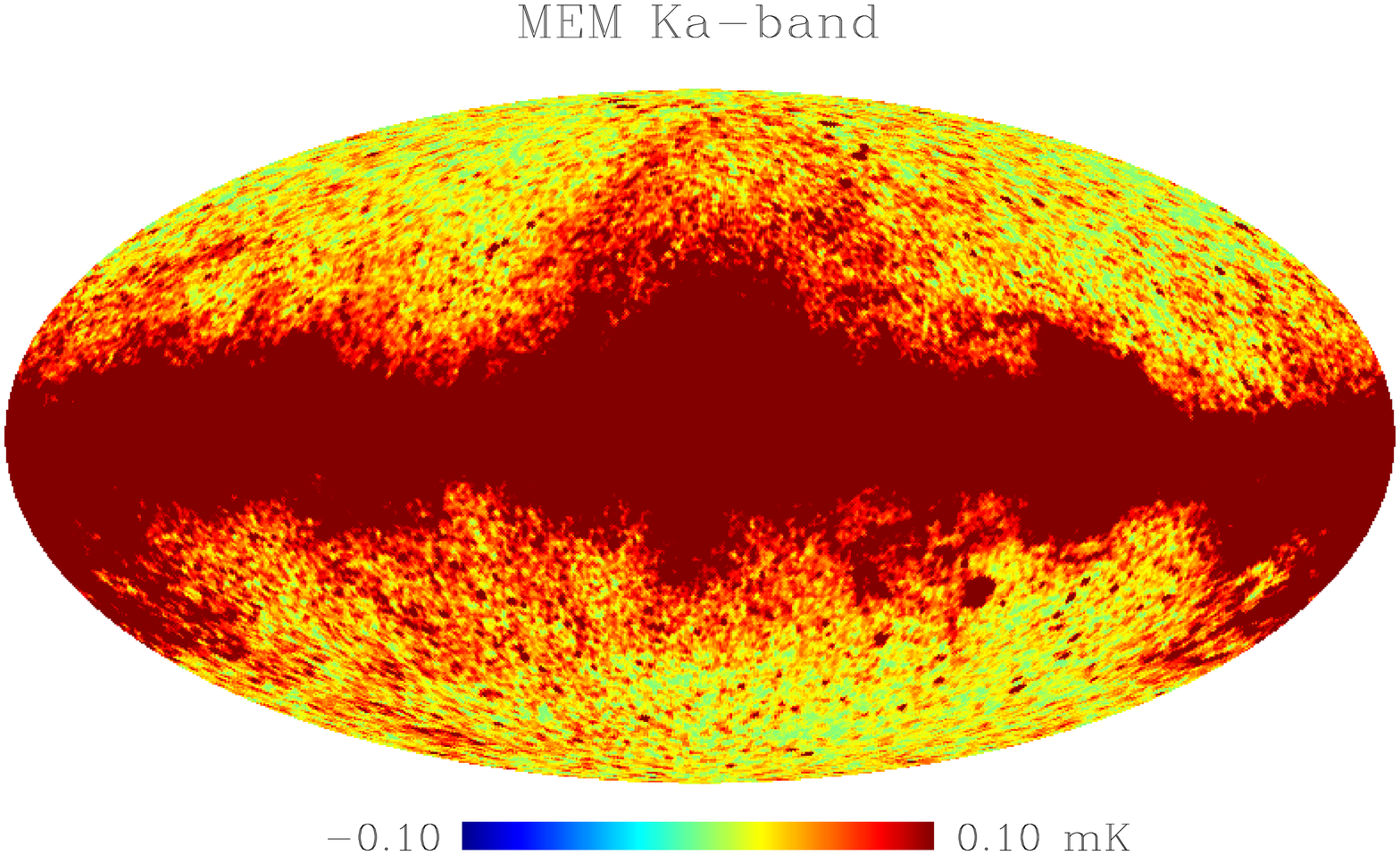,width=0.2\linewidth,angle=90,clip=} &
\epsfig{file=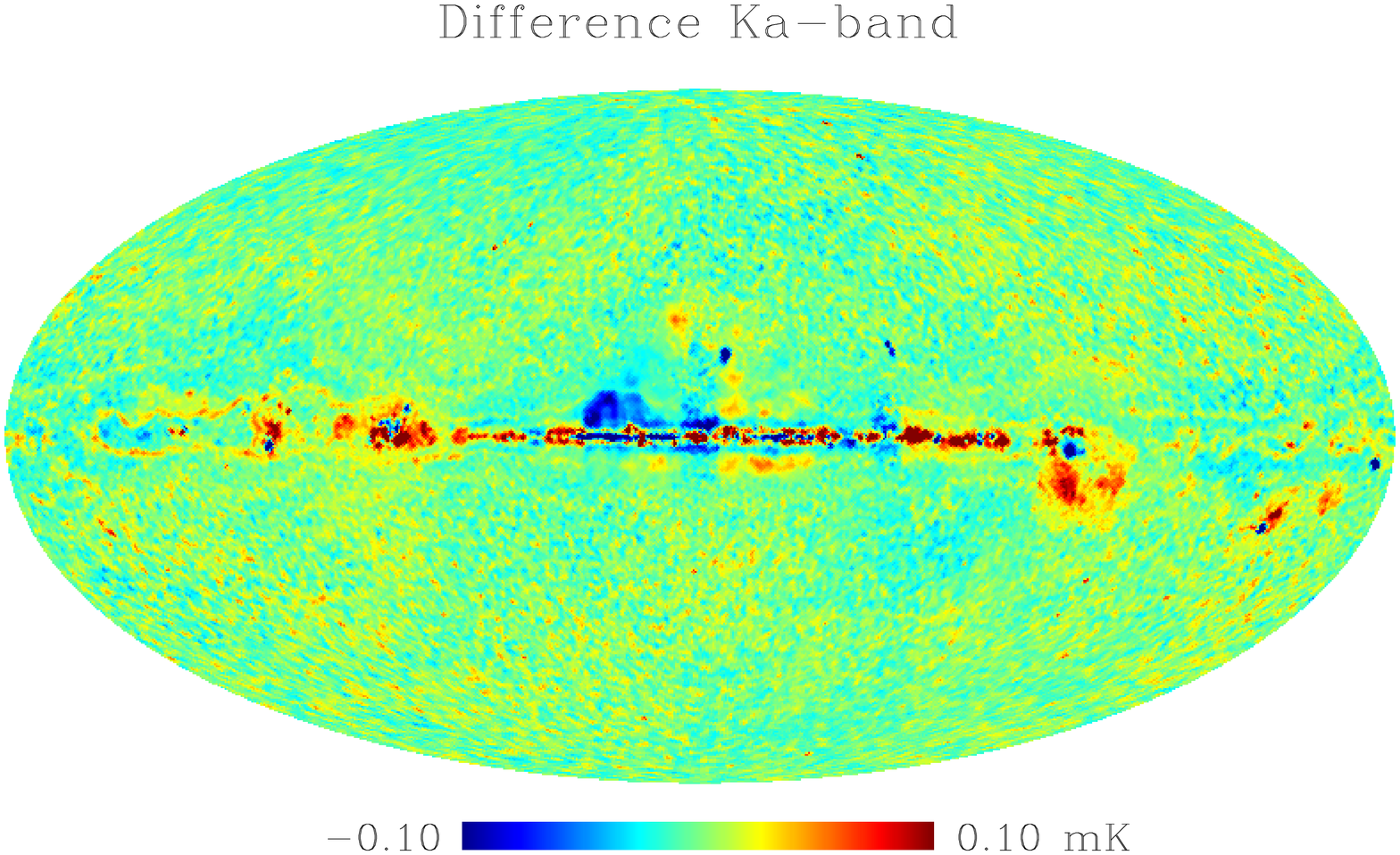,width=0.2\linewidth,angle=90,clip=} \\
\epsfig{file=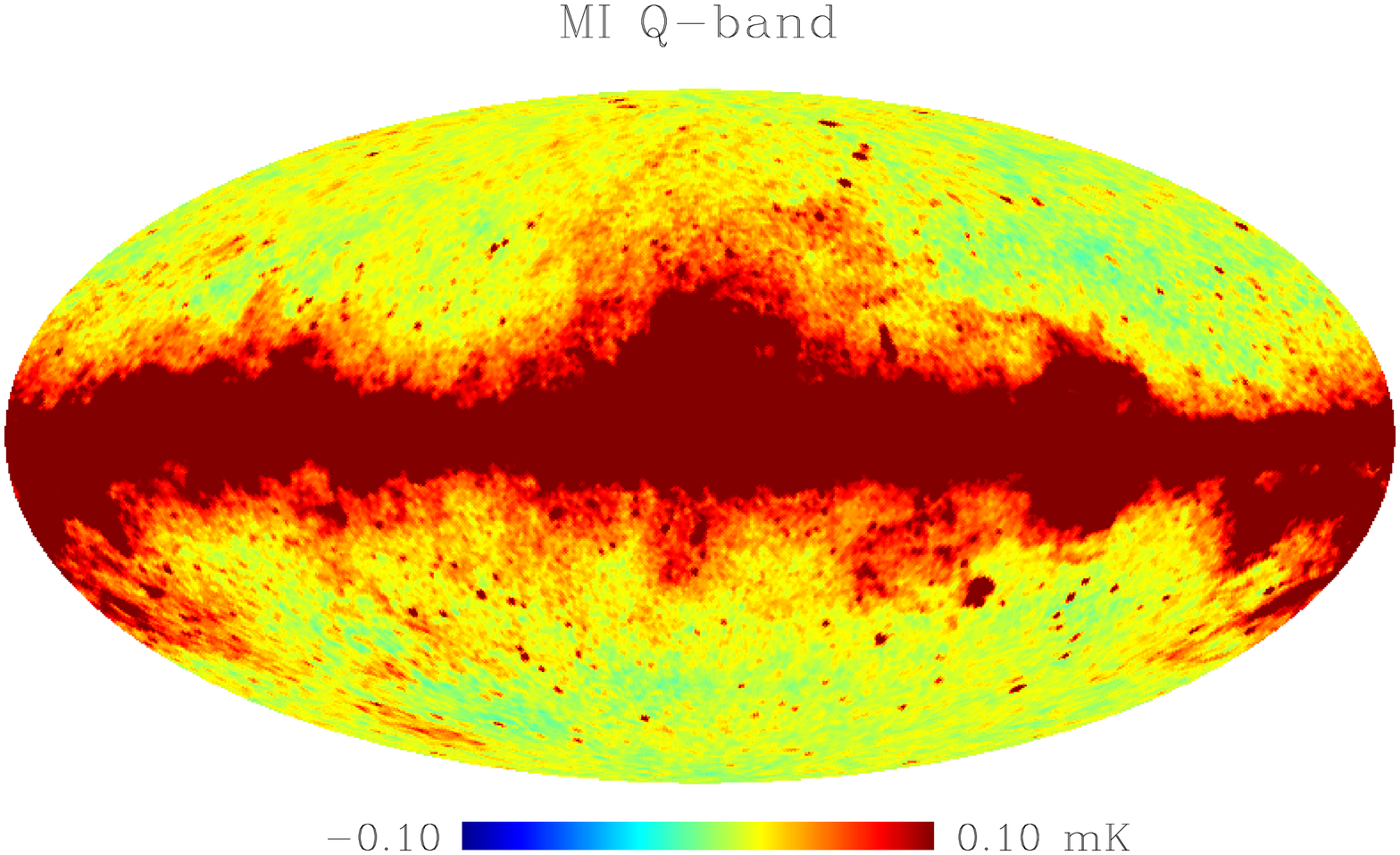,width=0.2\linewidth,angle=90,clip=} &
\epsfig{file=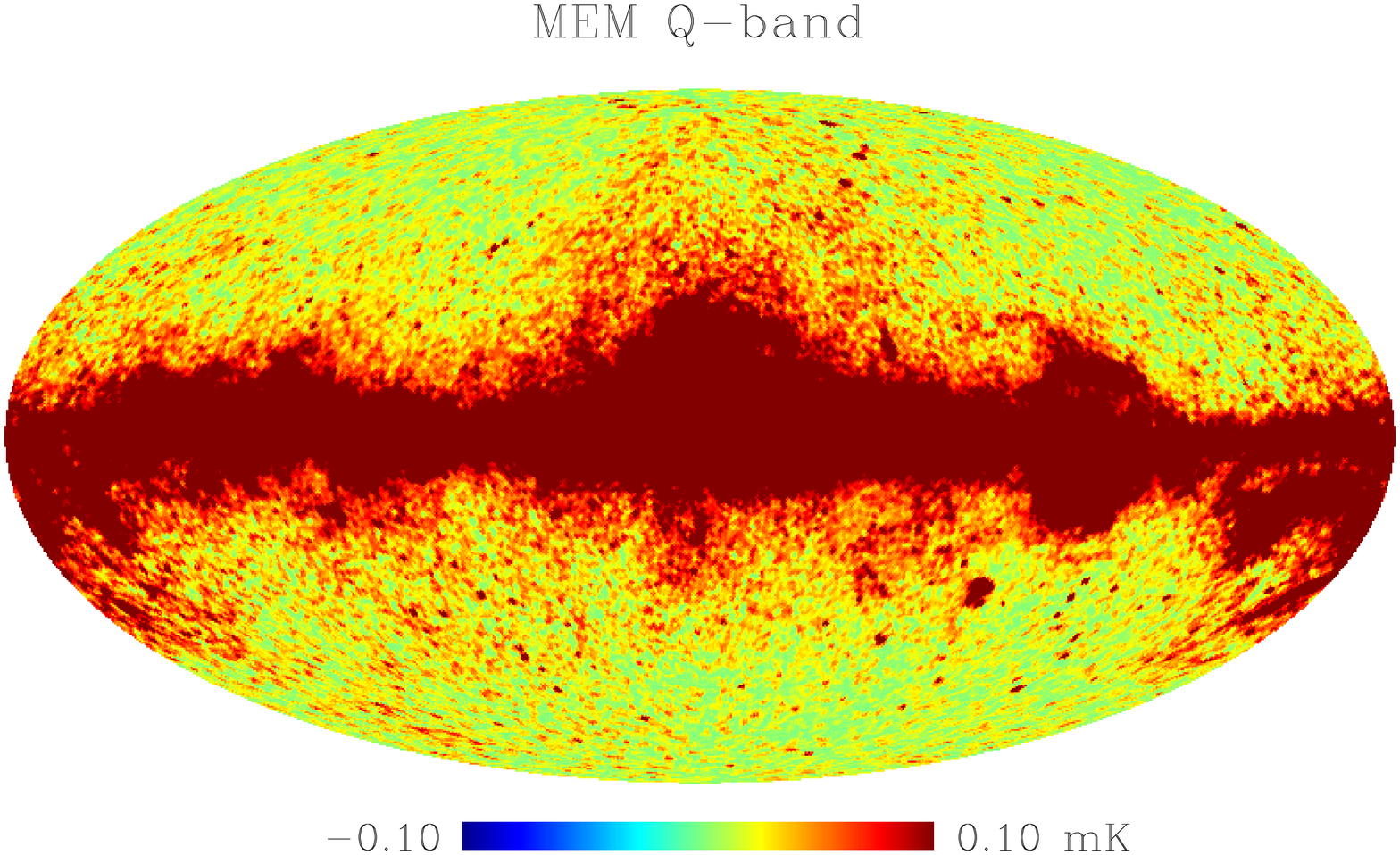,width=0.2\linewidth,angle=90,clip=} &
\epsfig{file=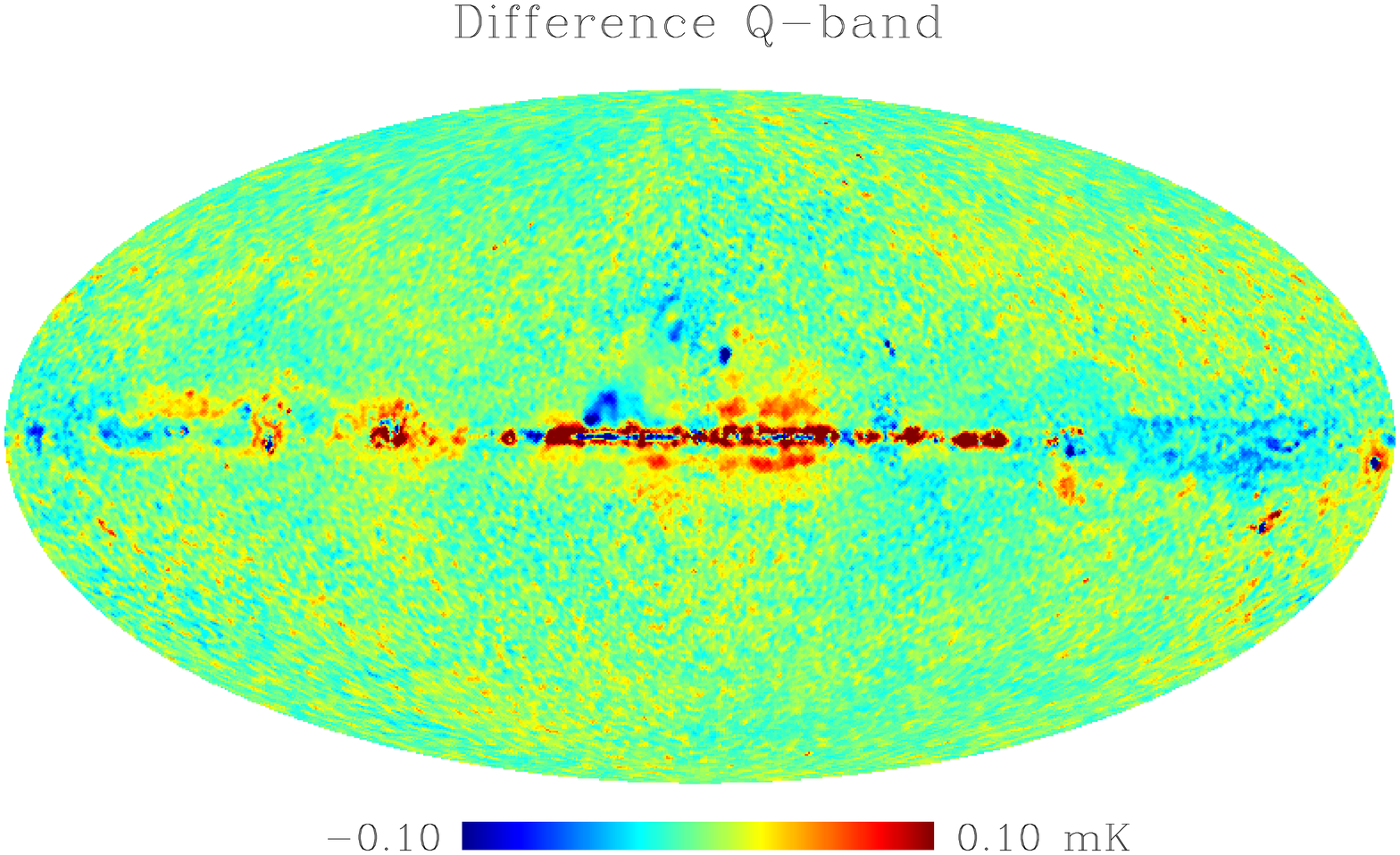,width=0.2\linewidth,angle=90,clip=} \\
\epsfig{file=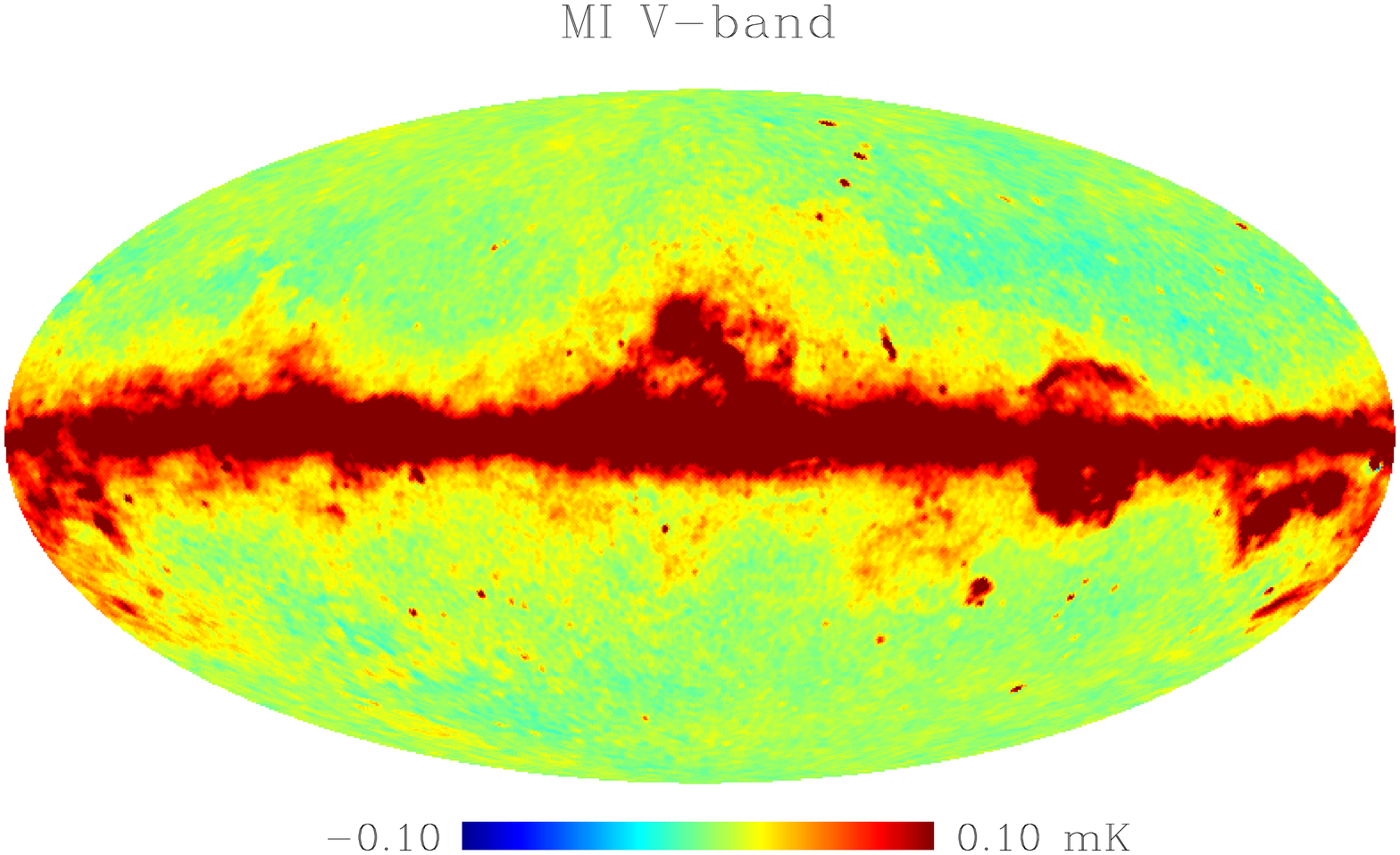,width=0.2\linewidth,angle=90,clip=} &
\epsfig{file=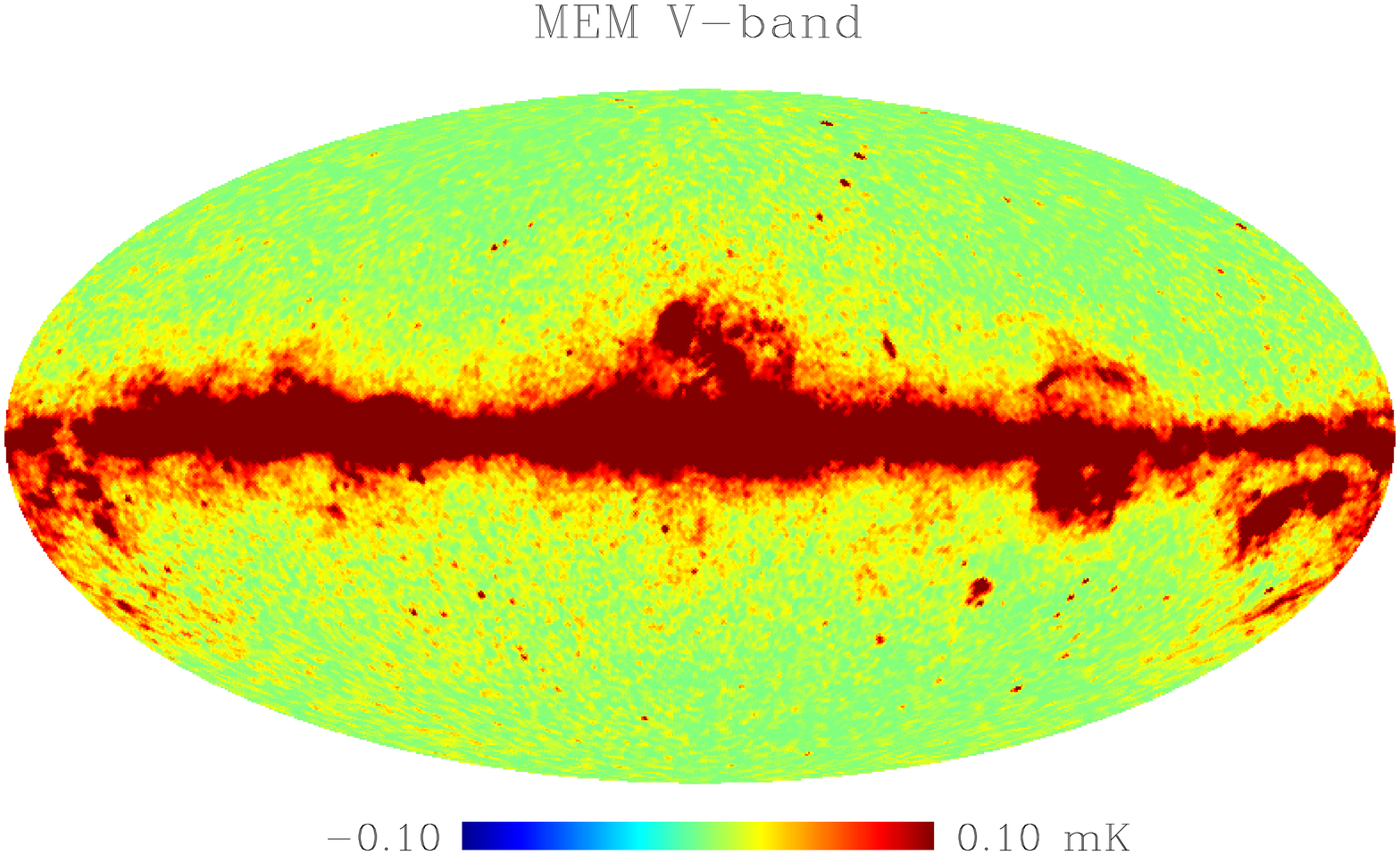,width=0.2\linewidth,angle=90,clip=} &
\epsfig{file=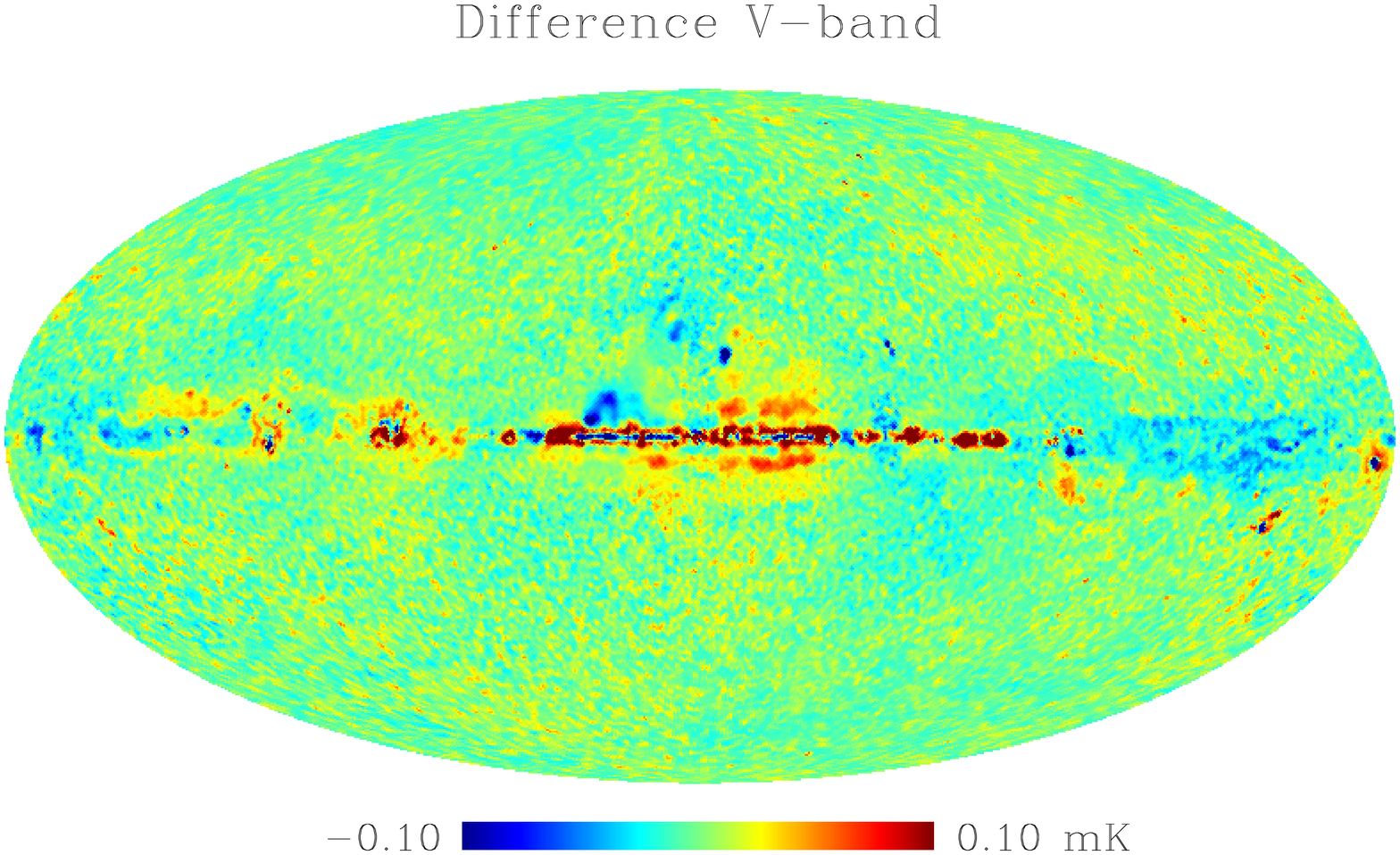,width=0.2\linewidth,angle=90,clip=}\\
\epsfig{file=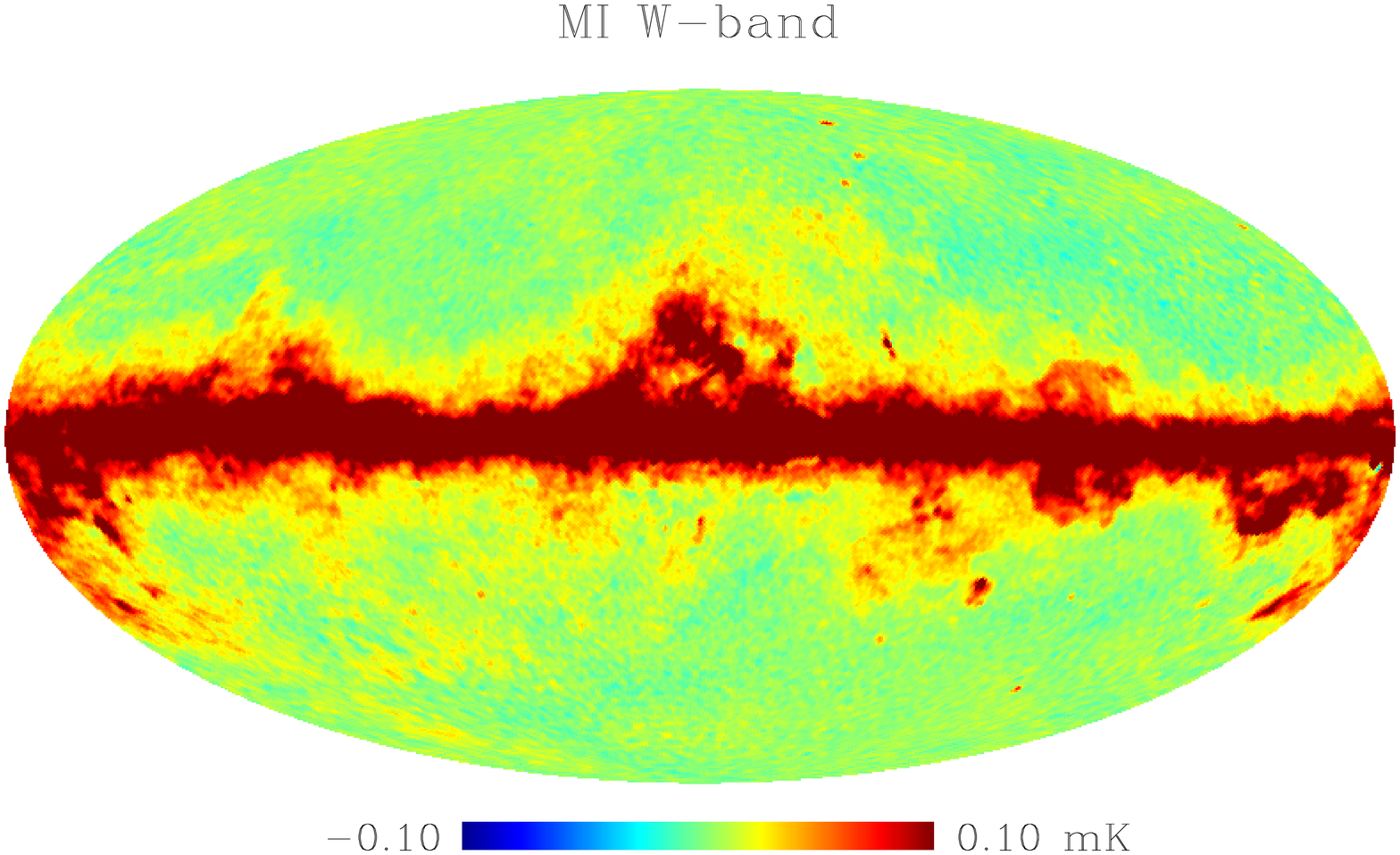,width=0.2\linewidth,angle=90,clip=} &
\epsfig{file=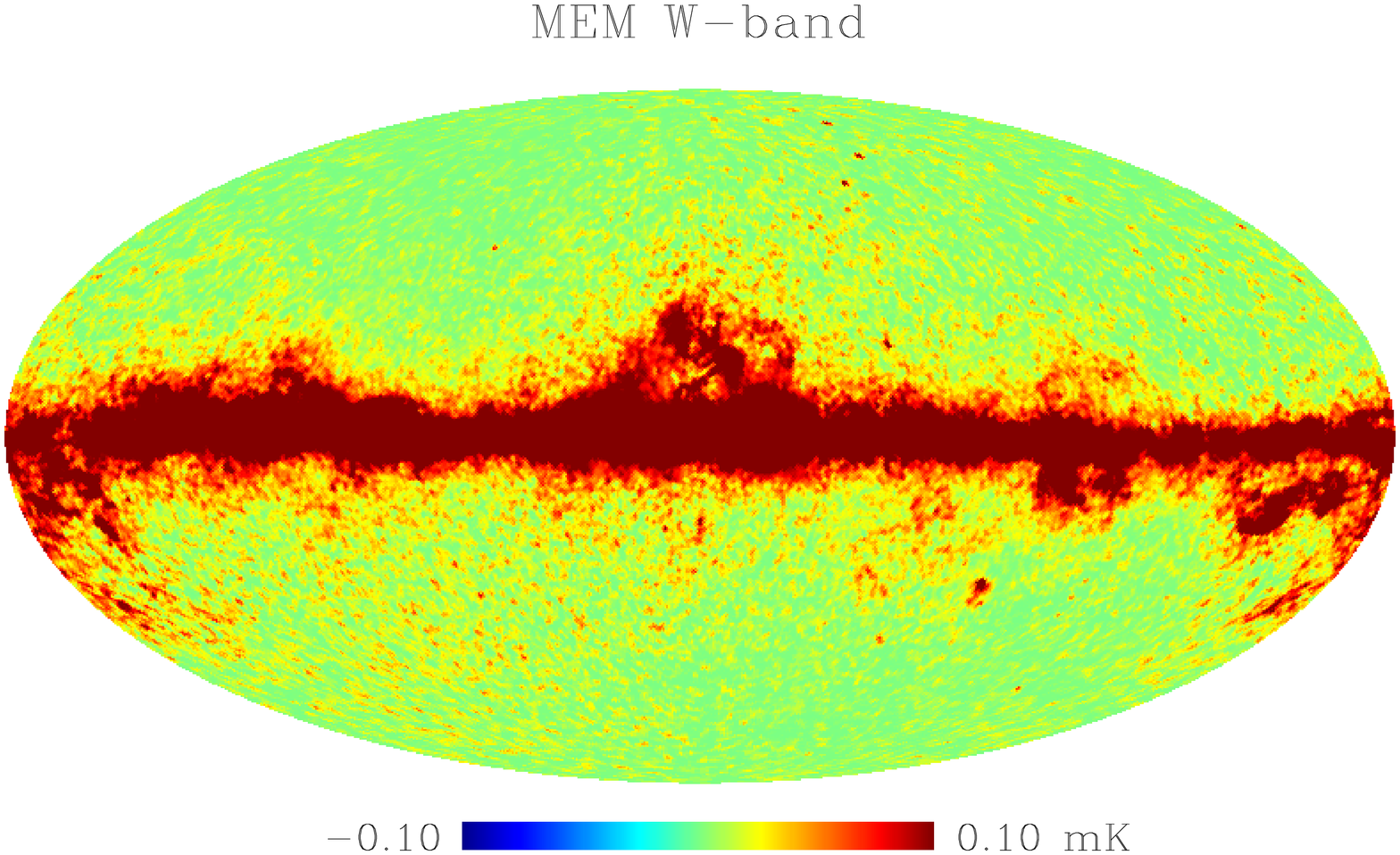,width=0.2\linewidth,angle=90,clip=} &
\epsfig{file=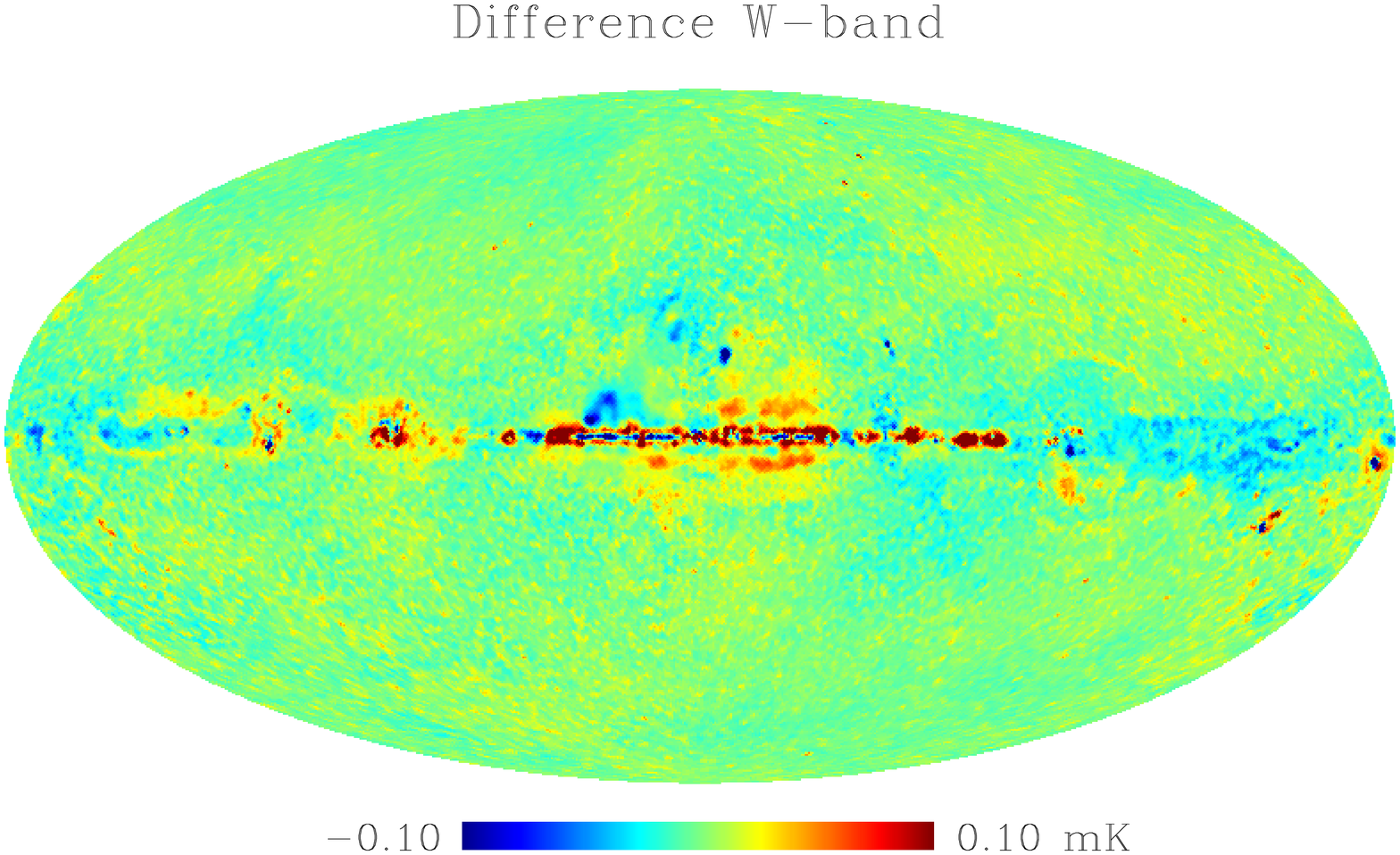,width=0.2\linewidth,angle=90,clip=}
\end{tabular}
\caption{The first column is the combined foreground maps for K, Ka, Q, V and W maps from top to bottom obtained by model independent analysis. The second column represents the combined foregrounds maps for K, Ka, Q, V and W bands obtained by MEM method and the third column is MEM minus model independent foreground maps. Model independent foreground maps for K and Ka are obtain from 3 channel clean maps whereas the model independent foreground maps for Q, V and W band are obtained from 4 channel clean maps.}
\end{figure}

\subsection{Verifying the Model Independent Method using Monte-Carlo Simulations}

We carry out a set of Monte-Carlo simulations to  estimate error on the rms foreground temperature obtained by the model independent analysis. A set of 1000 simulations of CMB maps are made using the HEALPix software for $N_{side}$=128 with proper beam-width and noise properties for each frequency channel of WMAP 5yr. We choose to downgrade WMAP maps to $N_{side}$=128 for our 1000 simulations because foregrounds are important at low multipoles. We use the publicly available Planck Sky Model (PSM) version 1.1, to generate the diffuse foreground maps. We estimate 48 (from 4 channel cleaning) and 24 (from 3 channel cleaning) cleaned maps from each realizations as	discussed in section II. We subtract them from the input maps and then follow the steps as discussed in section III. Finally the rms temperature of foreground from 1000 realizations are recovered and compared with the input values. The mean and variance of recovered foreground rms values are given in table~\ref{table2}. For 1000 simulations, we get 72000 cleaned maps for Nside=128. The simulations shows that our model independent foreground analysis recovers the foreground r.m.s power very well.

\begin{table}[h!]
\begin{tabular}{|c|c|c|}
\hline
Frequency &  ${\Delta T}_{rms}$ (in $\mu$K)    &   ${\Delta T}_{rms}$ (in $\mu$K)   \\
\textit{(in GHz)} &  \textit{Input PSM Template}    &  \textit{Extracted using Our Analysis}   \\
\hline
23   & 1440.73 &  1440.64 $\pm$ 3.98 \\
33   &  628.51 &   627.67 $\pm$ 3.86  \\
41   &  396.01 &   396.29 $\pm$ 3.94  \\
61   &  206.85 &   207.23 $\pm$ 3.62   \\
94   &  181.71 &   182.27 $\pm$ 2.97  \\
\hline
\end{tabular}
\caption{Comparison of input foreground rms temperature (expressed in mK) with the recovered foreground rms temperature obtained by Monte-Carlo simulations.}
\label{table2}
\end{table}

While comparing the PSM recovered rms temperature of foregrounds from simulations with that obtained by MEM analysis and our analysis, we note that the PSM foreground templates significantly underestimate the level of galactic contamination. The PSM generated foreground power spectrum for K band is much lower than the MEM generated power spectrum. But the differences decrease with the decrease of level of foreground contamination and for W band it matches with the MEM templates. This result has also been noted by other groups~\cite{sara}. 

\section{Estimation of Synchrotron spectral index over different regions of sky.}
It is well established that spectral index of synchrotron emission varies significantly over the sky. We estimate the spectral index of synchrotron emission over different regions of the sky using the model independent foreground estimation. The regions are defined by the 192
coarser pixels of HEALPix Pixelization at Nside=4. Synchrotron spectral index can be calculated easily by knowing the frequency dependence of $\Delta T_{rms}^F$. By definition $\Delta T_{rms}^F$ is given by,
\begin{equation}
\Delta T_{rms}^F = \left \lbrace \frac{1}{N_{pix}} \sum_{i=1}^{N_{pix}} \langle \Delta T^F(\hat{n})\rangle^2 \right \rbrace^{\frac{1}{2}},
\end{equation}
where the index `$F$' represents the combined foreground emission coming from dust emission, free-free emission, synchrotron emission and $N_{pix}$ denotes the number of pixels at resolution Nside=4 in the given region of the map. $\Delta T_{rms}^F$ is generally expressed in terms of antenna temperature.

Using equation~\eqref{eq:c}, the cleaned map in real space can be written as a sum of CMB map and residual noise as,
\begin{equation}
 \Delta T^{clean} =  \Delta T^{C} + \Delta T^{RN}.
\end{equation}
Multiplying $\Delta T^{clean}$ with $1^0$ resolution beam $B(\hat{n}.\hat{n}')$ and subtracting from equation~\eqref{eq:a}, we get
\begin{align*}
 \Delta T^{R}(\hat{n}) &= \int (\Delta T^{F}(\hat{n}) + \Delta T^{RN}(\hat{n})) B(\hat{n}.\hat{n}')\,d\hat{n}' + \Delta T^{N}(\hat{n}) \\
&= \Delta T'^{F}(\hat{n}) + \Delta T^{N'}(\hat{n}),
\end{align*}
where, $\Delta T^{R}(\hat{n})$ denotes the map of foreground plus residual noise. The beam smoothed foreground map can be defined as,
\begin{equation*}
 \Delta T'^{F}(\hat{n}) = \int \Delta T^{F}(\hat{n}) B(\hat{n}.\hat{n}')\,d\hat{n}'.
\end{equation*}
We can define a quantity $\Delta T_{rms}^R$ as,
\begin{equation}
\label{eq:f}
(\Delta T_{rms}^R)^2 = \frac{1}{N_{pix}}\sum_{i=1}^{N_{pix}} \langle \Delta T^R_a(\hat{n})  \Delta T^R_b(\hat{n})\rangle,
\end{equation}
where the index `$a$' and `$b$' represents the two independent detectors whose noise are uncorrelated. Using the above relation, $(\Delta T_{rms}^R)^2$ can be written as,
\begin{align}
 (\Delta T_{rms}^R)^2 &= \frac{1}{N_{pix}}\sum_{i=1}^{N_{pix}} \langle \Delta T'^{F}_a(\hat{n}) \Delta T'^{F}_b(\hat {n})\rangle  + \langle \Delta T_{a}^{N'}(\hat {n}) \Delta T_{b}^{N'}(\hat {n})\rangle \notag\\
\label{eq:g}
&= {(\Delta T_{rms}^F)}^2 + {(\Delta T_{rms}^{N'})}^2. 
\end{align}
The assumption in the above calculation is that foreground and noise are uncorrelated. Since the noise for two independent detectors are uncorrelated, the second term vanishes and we retain only the foreground rms power. The foreground rms temperature for the frequency
channel with more than one DA's can be calculated as,
\begin{equation}
 \Delta T_{rms}^F = \Delta T_{rms}^R = \left \lbrace \frac{1}{N_{pix}}\sum_{i=1}^{N_{pix}} \langle \Delta T^R_a(\hat{n})  \Delta T^R_b(\hat{n})\rangle \right \rbrace^{\frac{1}{2}}.
\end{equation}
For Q to W band, where the number of DA's is ranging from two to four, the foreground rms temperature can be calculated using the above equation. For example to calculate the quantity $\Delta T_{rms}^F$ for Q-band, the steps are as follows:

\begin{itemize}
\item Smooth all the input WMAP DA maps and the cleaned maps to one degree beam resolution.
\item Subtract the cleaned map C1 obtained by 4 channel combinations from Q1 band and similarly subtract cleaned map C12 obtained by 4 channel combinations from Q2 band. The CMB subtracted maps are label as Di given in table~\ref{tab:2}.
\item Take the cross product of DO1(Q1-C1) with the D13(Q2-C12) over the region of sky defined by a single pixel of Nside=4 and take the sum over all the pixels excluding the pixels covered by WMAP5 point source mask as described in equation~\eqref{eq:f}. 
\item Repeat the above steps for 24 possible combinations produced by 4 channel cleaned maps and took the ensemble mean of it. As the noise properties of two difference maps D1 and D13 are uncorrelated, the second terms cancels out of equation~\eqref{eq:g}.
\end{itemize}

\begin{figure}[h!]
\begin{tabular}{cc}
\epsfig{file=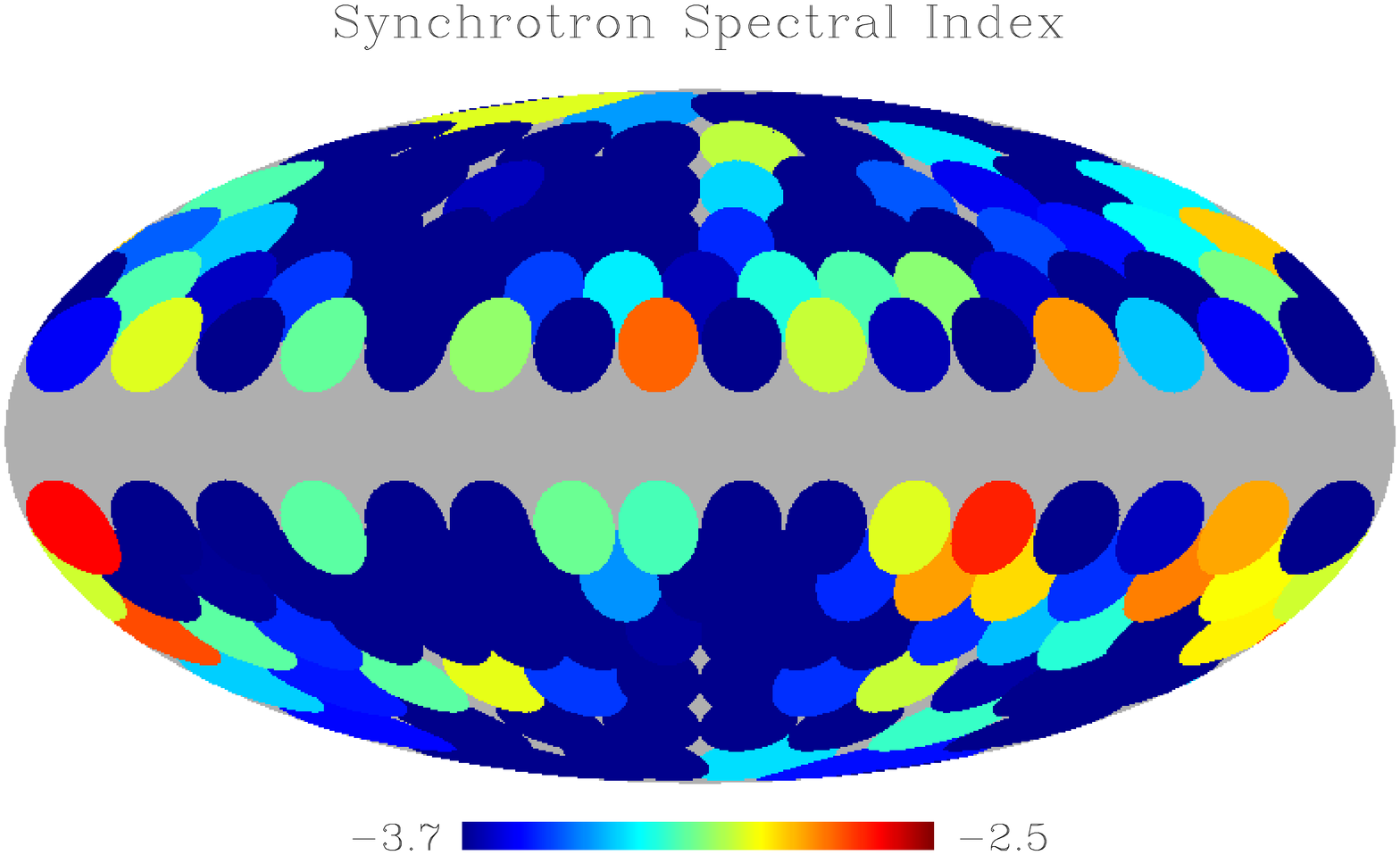,width=0.3\linewidth,angle=90,clip=}&
\epsfig{file=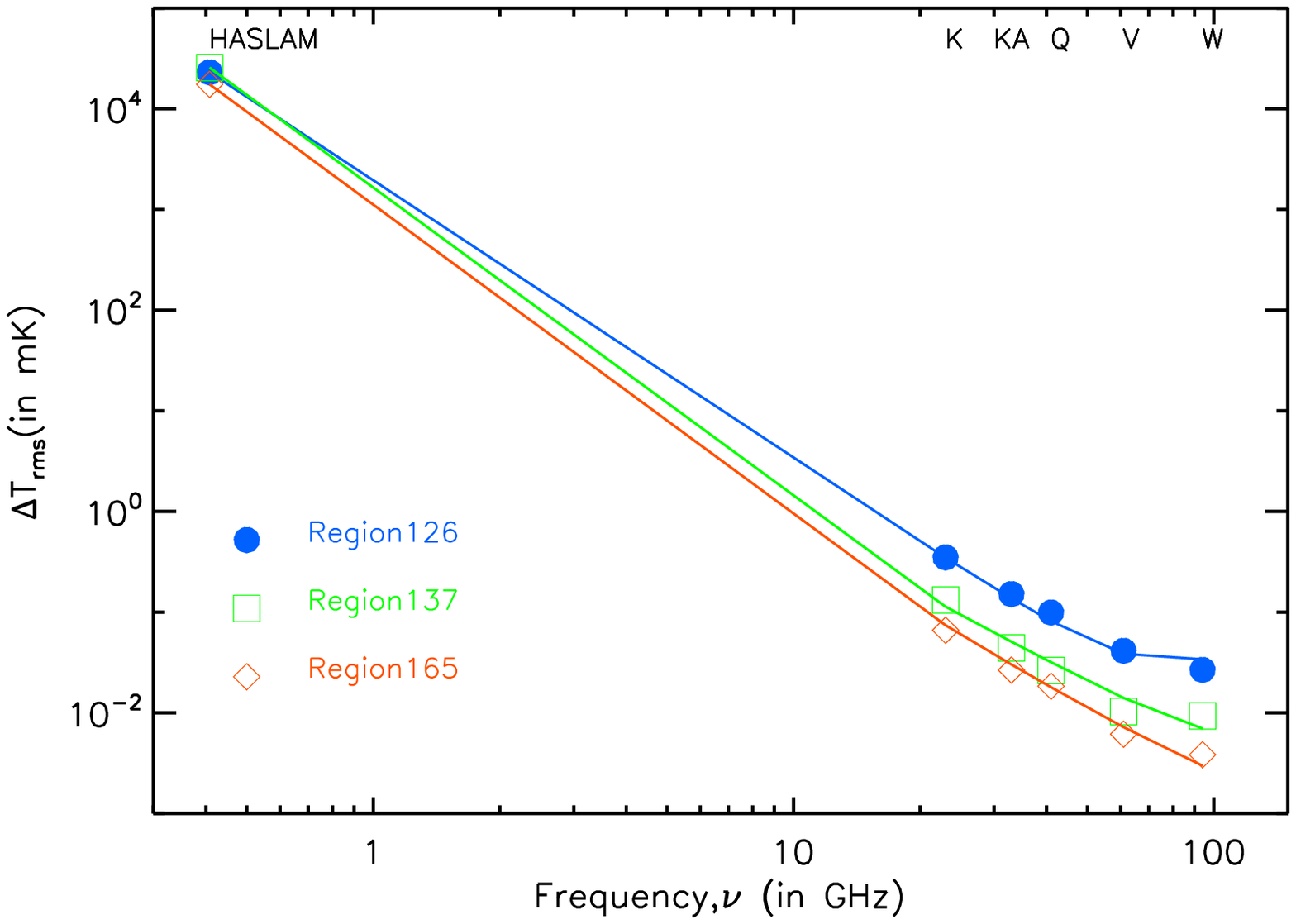,width=0.5\linewidth,angle=0,clip=}\\
\end{tabular}
\caption{Synchrotron spectral index variation over different positions of the sky obtained by model independent foreground analysis. The spectral index behaviour clearly shows that $\beta_s$ is -3.5 at high latitudes and -2.5 close to the galactic plane which is consistent with WMAP Team. For example the fit for 3 regions are shown in the plot. The region 126 (latitude range [$10^0 < \ell < 29.5^0] $, longitude range$ [ 248^0 < b <269^0] $, synchrotron spectral index$, [\beta_s=-2.82]$), region 137 ($-59.1^0 < \ell < -30.7^0$, $23.1^0 < b < 44.5^0, \beta_s =-3.70$), region 165 ($-53.3^0 <\ell < -30.3^0, 248^0 < b < 267^0, \beta_s=-3.33$).} 
\label{fig:3a}
\end{figure}

But for K and Ka band where only one DA is present, we subtract 3 channel cleaned maps C1 from K band and subtract 3 channel cleaned map C12 from K band and took the cross product over the region of sky to get rid of residual noise coming from model independent analysis of CMB
power spectrum. The equation~\eqref{eq:g} for the case of frequency channels with one DA becomes,
\begin{equation*}
 (\Delta T_{rms}^R)^2 = {(\Delta T_{rms}^F)}^2 + {(\Delta T_{rms}^{N})}^2,
\end{equation*} 
where $\Delta T_{rms}^{N}$ can be defined as,
\begin{equation*}
(\Delta T_{rms}^N)^2 = \frac{1}{N_{pix}}\sum_{i=1}^{N_{pix}} \langle \Delta T^N_a(\hat{n})  \Delta T^N_b(\hat{n})\rangle.
\end{equation*}
which has been estimated  using 1000 simulations of noise maps for K and Ka band smoothed to 1 degree beam resolution over the given region of the sky. The rms foreground temperature for K and Ka band can be calculated using the relation,
\begin{align}
 \Delta T_{rms}^F&= \left \lbrace \frac{1}{N_{pix}}\sum_{i=1}^{N_{pix}} \langle \Delta T^R_a(\hat{n})  \Delta T^R_b(\hat{n})\rangle\right \rbrace ^{\frac{1}{2}}- \Delta T_{rms}^N.
\end{align}
The $\Delta T_{rms}^F$ obtained is in terms of thermodynamic temperature which is converted to antenna temperature using the table~\ref{table1}. For each region, the frequency dependence of
$\Delta T_{rms}^F$ is fitted using the relation,
\begin{equation*}
\Delta T_{rms} = A_s\nu^{\beta_s}+A_f\nu^{\beta_f}+ A_d\nu^{\beta_d},
\end{equation*}
where $\beta_d=1.8$, $\beta_f=-2.14$ are taken as a constant parameters.  We fit $\Delta T_{rms}^F$ vs frequency to calculate the synchrotron spectral index at each region of the sky. To break the degeneracy, we use the Haslam map as a input template to increase the degree of freedom while computing the synchrotron spectral index. The rms noise level of Haslam map is 0.5K at high latitudes and 0.7K at low latitudes. Since the Haslam map isn't corrected for noise so the resulting spectral index variation puts an upper limit on the value of average $\beta_s$. The spectral index variation at each pixel is shown in figure(\ref{fig:3a}) as a sky map where we replace each HEALPix pixel with a colored circle which fits inside the pixel. We exclude the galactic region between $\pm$ $5^0$ latitude in our analysis.

\section{Discussions}

The estimation of foreground power spectrum from WMAP is carried out in a self contained method without using any extra information at any other frequencies other than WMAP frequencies. This work can be considered as an comprehensive approach that estimates both the CMB power spectrum and the foreground power spectrum simultaneously in a model independent approach. The method described in this paper is unbiased, we established through Monte-Carlo Simulations. Table~\ref{table2} shows that the recovered foreground rms power is very close to input
foreground rms power we put using the PSM template. Interestingly we find that MEM method overestimates the foreground power close to galactic plane and underestimates the foreground power at high latitudes relative to our estimates. But for the full sky, the mean rms foreground power using model independent method is close to MEM method. We find that the average synchrotron spectral index from K to W band is $\beta_s=-2.6$ over the full sky. The behaviour of synchrotron spectral index of $\beta_s=-3.5$ at high latitudes and $\beta_s=-2.5$ close to the galactic plane is consistent with WMAP Team. The advantage of this method is the cross-correlations takes care of the residual noise which remains after cleaning the map. For the upcoming PLANCK mission, this method of foreground power estimation will be even more promising since there are huge number of cross-combinations available due to large number of detectors of it and greater frequency coverage of PLANCK.

\section{Acknowledgement}

Computations were carried out on Cetus, the high performance computation facility of IUCAA. Some of the results in this paper have used the HEALPix \cite{healpix} Package. We acknowledge the use of the Legacy Archive for Microwave Background Data Analysis
(LAMBDA). Support for LAMBDA is provided by the NASA office of Space Science. We acknowledge the use of version 1.1 of the Planck reference sky model, prepared by the working group 2 and available at \href{www.planck.fr/heading79.html}{www.planck.fr/heading79.html}. A portion of the research described in this paper was carried out at the Jet Propulsion Laboratory, California Institute of Technology, under a contract with the National Aeronautics and Space Administration. TG thanks Council of Scientific Research and Industrial Research, India for financial support.

\end{document}